\documentclass[structabstract]{aa}  

\usepackage{hyperref}
\usepackage{natbib}
 \bibpunct{(}{)}{;}{a}{}{,} 
 
\usepackage{graphicx}

\usepackage{txfonts}
\def\kms{km s$^{-1}$\space}
\def\kmsno{km s$^{-1}$}
\def\micron{$\mu$m\space}

\def\micronno{$\mu$m}
\def\arcsecno{$^{\prime\prime}$}
\def\arcsec{$^{\prime\prime}$\space}

\def\arcmin{$^{\prime}$\space}
\def\deg{$^{\circ}$\space}

\def\degno{$^{\circ}$}
\def\h2{H$_2$}
\def\Ms{M$_\odot$}

\def\cii{[C\,{\sc ii}]\space}

\def\ciino{[C\,{\sc ii}]}

\def\ci{[C\,{\sc i}]\space}
\def\cino{[C\,{\sc i}]}
\def\hi{H\,{\sc i}\space}

\def\hii{H\,{\sc ii}\space}

\def\12co{$^{12}$CO}
\def\13co{$^{13}$CO}
\def\c18o{C$^{18}$O}

\def\nii{[N\,{\sc ii}]\space}
\def\niino{[N\,{\sc ii}]}

\def\C+{C$^+$}
\def\h2{H$_2$}
\def\cm3{cm$^{-3}$}
\def\cm3s{cm$^{-3}$\space}
\def\cm2{cm$^{-2}$}
\def\cm2s{cm$^{-2}$\space}

\begin{document}

   \title{Kinematics and properties of the Central Molecular Zone as probed with \cii}
      
\titlerunning{The Central Molecular Zone as probed with \cii}

\authorrunning{W. D. Langer et al.}

   \author{W. D. Langer
             \inst{1},
 	 T. Velusamy
                  \inst{1}, 
            M. R. Morris
            \inst{2} ,
        	 P. F. Goldsmith
	 \inst{1}, 
	              \and	   
              J. L. Pineda
              	 \inst{1}     
                        }
      
         
   \institute{Jet Propulsion Laboratory, California Institute of Technology,
              4800 Oak Grove Drive, Pasadena, CA 91109, USA\\
              \email{William.Langer@jpl.nasa.gov}
	 \and                        
                  Department of Physics and Astronomy, UCLA, 
			403 Portola Plaza, Los Angeles, CA 90095-1547, USA
             }

   \date{Received 8 August 2016; Accepted 12 December 2016}

 

\abstract
{The Galactic Central Molecular Zone (CMZ)  is a region containing massive and dense molecular clouds, with dynamics driven by a variety of energy sources including a massive black hole. It is thus the nearest template for understanding  physical processes in extragalactic nuclei.  The CMZ's neutral interstellar gas has been mapped spectrally in many neutral atomic and molecular gas tracers, but the ionized and CO-dark \h2 regions are less well traced spectroscopically.}  
 {To identify features of the UV irradiated neutral gas, photon dominated regions (PDRs) and CO-dark H$_2$, and highly ionized gas  in the CMZ as traced by the fine structure line of \C+ at 158  \micronno, \ciino, and to characterize their properties.}  
 {We observed the \cii 158 \micron  fine structure line with high spectral resolution using \textit{Herschel} HIFI with two perpendicular On-the-Fly strip scans, along $l$ = --0\fdg 8 to +0\fdg 8 and $b$ = --0\fdg 8 to +0\fdg 8, both centered on ($l$,$b$) = (0\degno,0\degno). We analyze the spatial--velocity distribution of the \cii data, compare them to those of \ci and CO, and to dust continuum maps, in order to determine the properties and distribution of the ionized and neutral gas and its dynamics within the CMZ. } 
 {The longitude-- and latitude--velocity maps of \cii trace portions of the  orbiting open gas streams of dense molecular clouds, the cloud G0.253+0.016, also known as the Brick, the Arched Filaments, the ionized gas within several  pc of Sgr A and  Sgr B2, and  the Warm Dust Bubble.  We use the \cii data to determine the physical and dynamical properties of these CMZ features.}  
 {The bright far--IR 158 \micron \C+ line, \ciino, when observed with high spatial and spectral resolution, traces a wide range of emission features in the CMZ.  The \cii emission arises primarily from dense PDRs and highly ionized gas, and is an important tracer of the kinematics and physical conditions of this gas.  }

{} 

\maketitle



\section{Introduction}
\label{sec:introduction}

The Galaxy's Central Molecular Zone (CMZ) is a roughly 400 pc $\times$ 100 pc region stretching from $l\sim$ --1\fdg 0 to $\sim$ +1\fdg 5 that is a significantly different environment than the Galactic disk.  Among many prominent  features in the CMZ are a massive, $\sim$4$\times$10$^6$ \Ms, black hole, the Galactic Center Bubble containing the Arches and Quintuplet clusters \citep{morris1996,Molinari2014}, the Radio Arc and Arched Filaments \citep{YusefZadeh1984,Serabyn1987,YusefZadeh1987a}, and gas streams of dense molecular clouds orbiting the Galactic Center at a radius of 100 to 120 pc \citep{Tsuboi1999,Molinari2011,Jones2012,Kruijssen2015,Henshaw2016b,Henshaw2016a}. The CMZ molecular clouds contain about 10\% of the Galaxy's molecular gas, have high average densities ($>$10$^4$ cm$^{-3}$), reside in a high thermal pressure environment, are relatively densely packed \cite[][]{morris1996,Ferriere2007}, and have regions with an intense flux of stellar Far-UV radiation as well as X-rays \cite[see review by][]{Ponti2013}. There is ample evidence of large-scale energetic gas motions in the CMZ, as indicated by the  line widths of CO in Giant Molecular Clouds (GMCs) of 10 to 20 \kmsno, compared to a few \kms in the Galactic disk  \cite[cf.][]{Oka1998a,Dame2001}, and by the flows of material within and across this region \cite[c.f.][]{morris1996,Morris1997,enokiya2014}.  The CMZ produces $\sim$10\% of the Galaxy's infrared luminosity, yet has a relatively low star formation rate, considering the mass of dense molecular gas there \citep{Taylor1993,Longmore2013a}.  

The CMZ's interstellar medium has been fully or partially mapped spectrally in over twenty gas tracers, including several isotopologues \citep{Morris1997,Jones2012}.  The  vast majority trace the neutral gas; these include the rotational transitions of CO isotopologues \cite[e.g.][]{Oka1998a,Oka2011,Oka2012,enokiya2014}, CS \citep{Tsuboi1999,Jones2012}, HCN \citep{Jackson1996,Jones2012}, N$_2$H$^+$ \citep{Jones2012}, H$_2$CO \citep{Ginsburg2016}, CH$_3$CCH \citep{Jones2012}, as well as NH$_3$ \cite[][]{Mills2013} and OH \citep{YusefZadeh1999}, and the lower fine structure emission line of neutral carbon, \ci \cite[][]{Martin2004}. However, these species do not trace major components of the inner Galaxy, where the gas is highly ionized  or where the UV irradiated neutral gas is weakly ionized.  These regions include the Photon Dominated Regions (PDRs), the ionized boundary layers (IBL) surrounding UV-irradiated molecular clouds, \hii regions, CO-dark \h2 clouds, and the warm and diffuse ionized gas. Less is known about these components due to the difficulty of observing key tracers of the ionized gas,  the fine structure lines of \C+ and N$^+$, from the ground. In this paper we present new results on the properties of the ionized and neutral gas  in the CMZ based on strip scans using the spectrally resolved fine structure line of ionized carbon.  

Prior to the launch of the  {\it Herschel Space Observatory}\footnote{{\it Herschel} is an ESA space observatory with science instruments provided by European-led Principal Investigator consortia and with important participation from NASA.}  \citep{pilbratt2010} the CMZ had been mapped in the fine structure line of C$^+$ at 158 \micronno, \ciino, with COBE FIRAS \citep{Bennett1994} and the Balloon-borne Infrared Carbon Explorer (BICE) \citep{Nakagawa1998}, but with its emission spectrally unresolved.  However, with the Heterodyne Instrument in the Far-Infrared (HIFI) \citep{degraauw2010} it became possible  to map spectrally resolved \cii in the CMZ.  The {\it Herschel} HEXGAL programme \citep{Guesten2007} made a well sampled HIFI  \cii On-the-Fly (OTF) map of a small region around Sgr A \citep{Garcia2015,Garcia2016} consisting of two slightly offset rectangles totaling $\sim$0.1 square degrees in area. As part of the {\it Herschel} GOT C+ programme \citep{Langer2010}, \cii was observed with HIFI along two spectrally resolved strip scans traversing 1\fdg 6 in longitude and latitude through ($l$,$b$) = (0\degno,0\degno).  The purpose of this paper is to gain insight on the nature of the CMZ as traced by these two \cii strip scans, which extend about half the length and the full width of the CMZ. We limit our discussion to describing the regions and features detected in \cii emission, their kinematics,  and estimates of the properties of the \C+ gas, as detailed information about the structure of  these features is not possible without spectrally resolved ($l$--$b$) maps well sampled in both longitude and latitude.   

While {\it Herschel} HIFI provides the necessary spectrally resolved instrument to study \ciino, the small beam size at 1.9 THz, 12\arcsecno, and single pixel made it unsuited to generate large-scale fully sampled maps, such as of the Galactic disk, CMZ, or external galaxies.  However, {\it Herschel} HIFI was able to provide well sampled detailed maps of small regions  or   a coarse picture of \cii emission across the Galaxy by conducting a large-scale sparse survey of spectrally resolved \ciino.  

The {\it Herschel} Open Time Key Programme, Galactic Observations of Terahertz C+ \cite[GOT C+;][]{Langer2010,Pineda2013} had two main subprogrammes: 1) a Galactic disk sparse pointed survey of \ciino; and, 2) a Central Molecular Zone OTF strip scan survey of \ciino.  In subprogramme (1) the GOT C+ survey characterized the entire Galactic disk by making a sparse sample covering 360\deg in the plane. (The GOT C+  data sets  are available as a \href{http://www.cosmos.esa.int/web/herschel/user-provided-data-products}{{\it Herschel} User Provided Data Product} under KPOT\_wlanger\_1). The disk survey contains $\sim$500 lines of sight of spectrally resolved \cii emission throughout the Galactic Disk ($l$=0\deg to 360\deg and $|b|$$\le$1\degno). The sampling in Galactic longitude was non-uniform in order to weight as best as possible a uniform angular volume across the disk, with an emphasis on the important inner longitude range $|l|$$\le$90\degno.  By analyzing a large sample of spectrally resolved components dispersed throughout the Galaxy, rather than large-scale maps of a few clouds, this survey allows a statistical approach to characterizing the ISM in \cii and the results have been reported in a series of papers \cite{Langer2010,Pineda2010,Velusamy2010,Velusamy2012,Pineda2013,Langer2014,Pineda2014,Velusamy2014,Velusamy2015}.  Here we present the results of the GOT C+  \cii strip scans of the CMZ,  which pass through, or close to, several features, including Sgr A, G0.253+0.016 (the Brick), Sgr B2, the Galactic Center Warm Dust Bubble, the Radio Arc, the Arched Filaments, and the open  streams of gas orbiting the Galactic Center.    

This paper is organized as follows.  In Section 2 we describe the data reduction, while Section 3 presents the distribution of the \cii traced gas compared to features in the CMZ.  Section 4 discusses the properties of several features traced by \cii in the CMZ, and Section 5 summarizes the results.



\section{\cii Observations and Data Reduction}
\label{sec:observations}

We observed the ionized carbon (C$^+$) $^2$P$_{3/2}$ -- $^2$P$_{1/2}$  fine structure line, \ciino, at 1900.5369 GHz ($\lambda  \sim$157.74 \micronno), using HIFI in an OTF mode to produce position--velocity maps of the Galactic center region. We observed two strip scans through $(l,b)$ = (0\degno,0\degno), one along longitudes from $l$=359\fdg20 to 0\fdg80 at latitude $b$=0\deg and the other along $b$= -0\fdg 80 to +0\fdg 80 at longitude $l$ = 0\degno;  details of the observations and data reduction are given below.    
 
\subsection{Observations}

All \cii spectral line mapping observations were made with the high spectral resolution HIFI instrument  \citep{degraauw2010} onboard {\it Herschel} \citep{pilbratt2010} in February and March 2011.  These \cii spectral line map scans used the HIFI band 7b  and the wide band spectrometer (WBS).  The longitude scan consisted of four OTF scans taken along Galactic longitude centered on $(l,b)$ = (0\fdg 6,0\fdg 0), (0\fdg 2,0\fdg 0), (359\fdg 8,0\fdg 0), and (359\fdg 4,0\fdg 0).  Similarly, the latitude scan consisted of four OTF scans taken in Galactic latitude centered on $(l,b)$ = (0\fdg 0,-0\fdg 6), (0\fdg 0,-0\fdg 2), (0\fdg 0,+0\fdg 2), and (0\fdg 0,+0\fdg 6).  Each OTF scans was 24 arcmin long, thus the four together cover a range of 1.6\degno.  The observing duration of each OTF scan was $\sim$2500 sec and the 24 arcmin-long scan data were read out every 40\arcsec (i.e. averaged over a 40\arcsec window).  

All HIFI OTF scans were made in the LOAD--CHOP mode using a reference off-source position about 2\deg away in latitude (this HIFI observing mode requires the reference position be within 2\deg of the target).  For all four $l$--scans and the two $b$--scans centered at $b \ge $0\fdg 0 we used the OFF ($l$,$b$) =(0\fdg0,1\fdg9) which is very clean of \cii emission at the sensitivity of the survey.  For the $b$--scans at $b<$0\fdg 0 we had to use a reference position closer to the plane at $(l,b)$ =  (0\fdg 0,0\fdg 5) which contains \cii emission. However, this reference position has been well observed in our HIFI Pointed observation programme using the reference sky position at (0\fdg 0,+1\fdg 9) so that the \cii emission spectrum at this position is well characterized and can be used as a  sky reference to correct for OFF source emission. 
Thus the mapping scheme used for the CMZ provides \cii spectral line map data all of which are effectively referenced back to a common position at $(l,b)$ = (0\fdg 0,1\fdg 9).  The GOT C+ survey data, as well as the OTF scan reference spectra extracted in HIPE-13, show little emission at a 3-$\sigma$ level of  $\sim$0.1K.
  
 Near the Galactic Center we expect \cii emission over a wide range of velocities ($>$200 \kmsno).  However, the useful instantaneous velocity range observable with HIFI band 7b is $\sim$160 \kmsno.  To accommodate this velocity range and  extract the spectral line data, each OTF scan was observed at two local oscillator (LO) frequencies with frequency offset of 0.9 GHz (corresponding spectra centered at $V_{LSR}$ =$-$80 \kms and +70 \kmsno).  
 
\subsection{Data reduction}
\label{sec:data}

We processed the OTF strip scan data and produced the $l$-- and $b$--scan position--velocity maps, ($l$--$V$) and ($b$--$V$), following the procedure discussed in \citet[][]{Velusamy2015}.  However, the results presented here were processed in HIPE-13, rather than HIPE-12 used in \cite{Velusamy2015}, taking advantage of the {\it hebCorrection} tool, which removes the HEB standing waves present in band 7b data.   {\it hebCorrection} is applied in HIPE-13 as part of the pipeline to extract spectra at the reference sky position.  From the Level 2 data, the \cii strip scans were made into ``spectral line cubes'' using the standard mapping scripts in HIPE.  Any residual  HEB and optical standing waves in the reprocessed Level 2 data  were minimized by  applying  {\it fitHifiFringe} to the ``gridded'' spectral data.  We took the additional precaution in {\it fitHifiFringe} of disabling  {\it DoAverage}  in order not to bias the spectral line window. 
      
The H-- and V--polarization data were processed separately and were combined only after applying {\it fitHifiFringe} to the gridded data. This approach minimizes the standing wave residues in the strip scans by taking into account   the standing wave differences between H-- and V--polarization.  In view of the fact that the spectrum of a single LO observation may not provide adequate baselines at both ends of  the spectral line profile in HIPE-13, we generated two sets of spectral cubes with and without baseline subtraction.  We then used the processed spectral line data cubes in two polarizations (H and V) and two LO settings to make ($l$--$V$) and ($b$--$V$) maps, by combining them after fitting baseline and  velocity offsets.  For $|l|$  $>$ 0\fdg4 and $|b|$ $>$ 0\fdg4 combining  the spectral cube data was straightforward as these spectra had clean baselines at both ends in one or both sets of the two LO settings.  

\begin{figure}
 \centering
             \includegraphics[width=9cm]{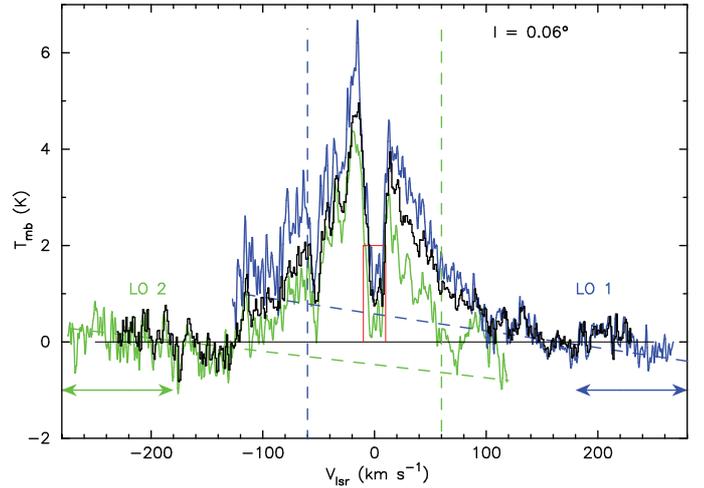}
      \caption{Example of combining the two LO spectra in OTF scan maps.  The spectra at $l$=0\fdg06 in the individual longitude scan maps consist of two local oscillator settings, LO 1 (blue) and LO 2 (green).  The final averaged spectrum is shown in black.  The double arrows mark the velocity range of a constant baseline offset used in the corresponding maps. The red box denotes the velocity range ($\pm$10 \kmsno) for matching the two spectra and the baseline offsets here were estimated using their averaged intensity.  The quasi--horizontal dashed lines represent a linear fit to the baselines used for the spectra for final averaging. The vertical dashed lines mark velocity limits above (green) and below (blue) where the data are unusable (near the edges of the band pass).  The spectra for velocities within $\pm$60 \kms were averaged with baseline corrections for V $<$ -60 and V$>$ +60 \kms for LO 2  and LO 1, respectively. Note that the absorption feature around 0 \kms is due to local gas. }
         \label{fig:fig1}      
 \end{figure}

However, for $|l|$$<$0\fdg4 and $|b|$ $<$0\fdg4 the spectra contained a long and clean baseline only at one end, with emission present at the opposite end.  In this case the spectra of the LO-pairs were combined, fixing the linear baseline at the cleaner end and matching the profiles at the overlapping velocities.  In Figure~\ref{fig:fig1} we show an example of this procedure for $l$=0\fdg06, where it can be seen that the matching produces  a good baseline correction.  For matching we use the region near $V_{LSR}$ $\sim$0 \kms  where \cii is absorbed by foreground emission.  To check this approach, we compared the resulting spectra for $|l|$ $>$0\fdg4, where the emission is narrow enough to fit within the band of a single LO, and the baselines can be set at both ends with no matching of profiles.  The resulting baselines and spectra fit both ways are in good agreement, thus lending confidence that the matching approach produces good baselines for  $|l|$$<$0\fdg4 and $|b|$ $<$0\fdg4. We find that this matching procedure does introduce a baseline uncertainty of $\sim$0.5 K and produces more uncertainty in the final spectra for the $l$ scan than for $b$ above and below the plane because these $b$ scans generally have narrower lines and, therefore longer and cleaner baselines.

At 1.9 THz the angular resolution of the {\it Herschel} telescope is 12\arcsecno, but  the \cii OTF observations  were averaged over a   40\arcsec window along the scan direction (i.e. read out at 40\arcsec intervals).  Such fast scanning broadens the effective beam size along the scan direction  \citep[][]{Mangum2007}.  Therefore all \cii maps have been restored with effective beam sizes corresponding to twice the sample averaging interval along the scan direction ($\sim$80\arcsecno). We  used the HIFI Wide Band Spectrometer (WBS) with a spectral resolution of 1.1 MHz (0.17 \kmsno) for all the scan maps. The final $l$--$V$ and $b$--$V$ maps presented here were smoothed to a velocity resolution of 2 \kmsno.  The rms in the main beam temperature, T$_{mb}$, is in the range 0.21 to 0.31 K for the  $l$--$V$ map and 0.18 to 0.22 K for the $b$--$V$ map, the smaller rms for the $b$--$V$ map is due to the cleaner baseline that results from having (generally) narrower line profiles along $b$ than along $l$.



\section{Results}
\label{sec:results}

In this section we present the \cii OTF strip scans and auxiliary spectral line maps of \ci and CO, and describe their morphological relationships to CMZ features.

\subsection{OTF maps}
\begin{figure*}[!ht]
 \centering
  \includegraphics[width=18cm]{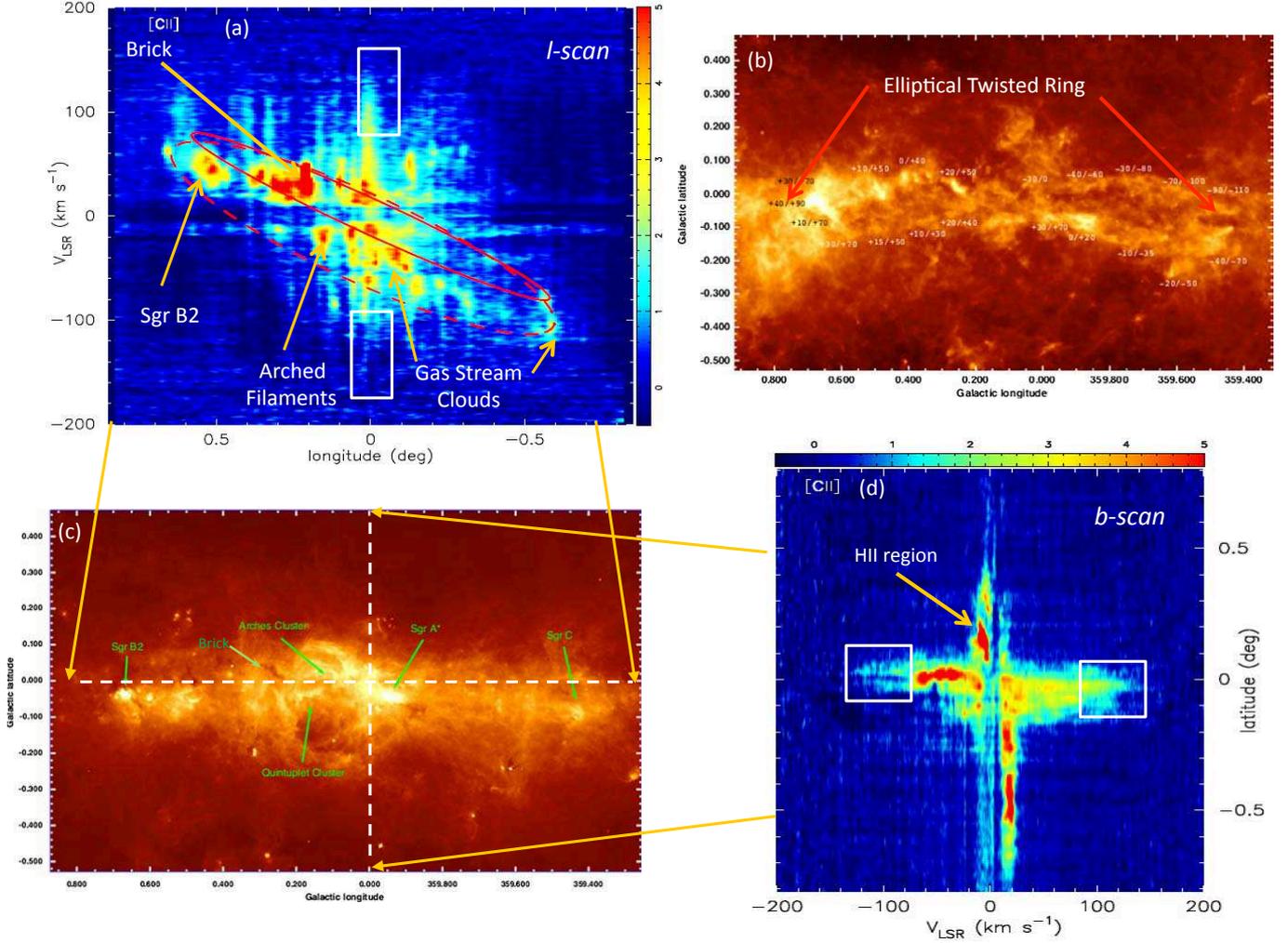}
         \caption{HIFI position--velocity \cii maps of the CMZ tracing several notable features, including the Brick, open orbit streams of molecular gas, Sgr B2, and the Arched Filament.   (a) The \cii  longitude--velocity map, $l$--$V$, covering 1\fdg 6 centered at (0\degno,0\degno).  The color bar indicates the \cii main beam temperature, T$_{mb}$(K). The red-solid ellipse is the solution at $z$=0\fdg 0 for the elliptical ring model from \cite{Molinari2011} in longitude--velocity space and the red-dashed ellipse is our fit to the \cii observations assuming an ellipse.  The vertical striations are artifacts of the baseline uncertainties (see Section~\ref{sec:observations}). (b) Atomic hydrogen column density map from \cite{Molinari2011} highlighting their elliptical twisted ring model.  The line-of-sight velocity along the ring was derived by \cite{Molinari2011} from CS observations and is marked along the ring (see the original figure).  (c) A {\it Herschel} PACS 70 \micron image of the CMZ from Figure 1 of \cite{Molinari2011}. The \cii OTF strip scans in $l$ and $b$ are indicated by white dashed lines. The labels in green are from \cite{Molinari2011}, except for the Brick (G0.253+0.016).  (d) The \cii latitude--velocity map, $b$--$V$, covering 1\fdg 6 centered at (0\degno,0\degno).  The colour bar indicates T$_{mb}$(\ciino) in degrees K. In both $l$--$V$ (panel a) and $b$--$V$ (panel d) we highlight the position--velocity extent of the high velocity gas near the center (white boxes). The arrow indicates a very bright \cii peak that is likely a foreground \hii source (see Section~\ref{sec:S17}).}
         \label{fig:fig2}            
  \end{figure*}

In Figure~\ref{fig:fig2} we show our HIFI GOT C+ position--velocity \cii  maps along with corresponding {\it Herschel} 70 \micron dust and \hi emission maps from \cite{Molinari2011}.  In Figure~\ref{fig:fig2}(a) we plot the longitude--velocity, $l$--$V$,  \cii emission and in (d) the latitude--velocity, $b$--$V$, emission.  The velocity dispersion along both the longitude and latitude strips is largest near $(l,b)$ = (0\degno,0\degno) and the highest detected velocities, as outlined by the white rectangles, reach about $\pm$150 \kmsno.   The locations in position--velocity space corresponding to a few features in the CMZ, including Sgr B2, the Brick, the Arched Filaments, and  the open orbit streams of clouds are marked on the $l$--$V$ plot. The low-intensity vertical striations in Figure~\ref{fig:fig2}(a) in the $l$--scan are artifacts of the baseline uncertainties.  These features are less prominent in the $b$--scan, as seen in Figure~\ref{fig:fig2}(b), because the baseline uncertainties are smaller (see Section~\ref{sec:data}). 
 
We have also highlighted the ellipse  (red solid line) in $l$--$V$ at $b$ = 0\deg corresponding to  the kinematics of a twisted elliptical ring  at $b$ = 0\deg which \cite{Molinari2011}  used to explain the distribution of CS velocities in Figure~\ref{fig:fig2}(b). The dashed (red) ellipse is our  fit to the \cii emission and will be discussed in Section~\ref{sec:Streams}.  Panel (b) shows an atomic hydrogen column density spatial map of the CMZ  with velocities from CS superimposed along portions of the elliptical ring from \cite{Molinari2011}.  Panel (c) shows the extent of the two OTF strip scans (dashed white lines) superimposed on a {\it Herschel} 70 \micron map from \cite{Molinari2011}.  Figure~\ref{fig:fig2}(d) plots the $b$--$V$ \cii emission at $l$ = 0\degno.   The velocity gradient is small in this map compared to that of the $l$--$V$ map.  
  
\begin{figure*}[!ht]
 \centering
            \includegraphics[width=18cm]{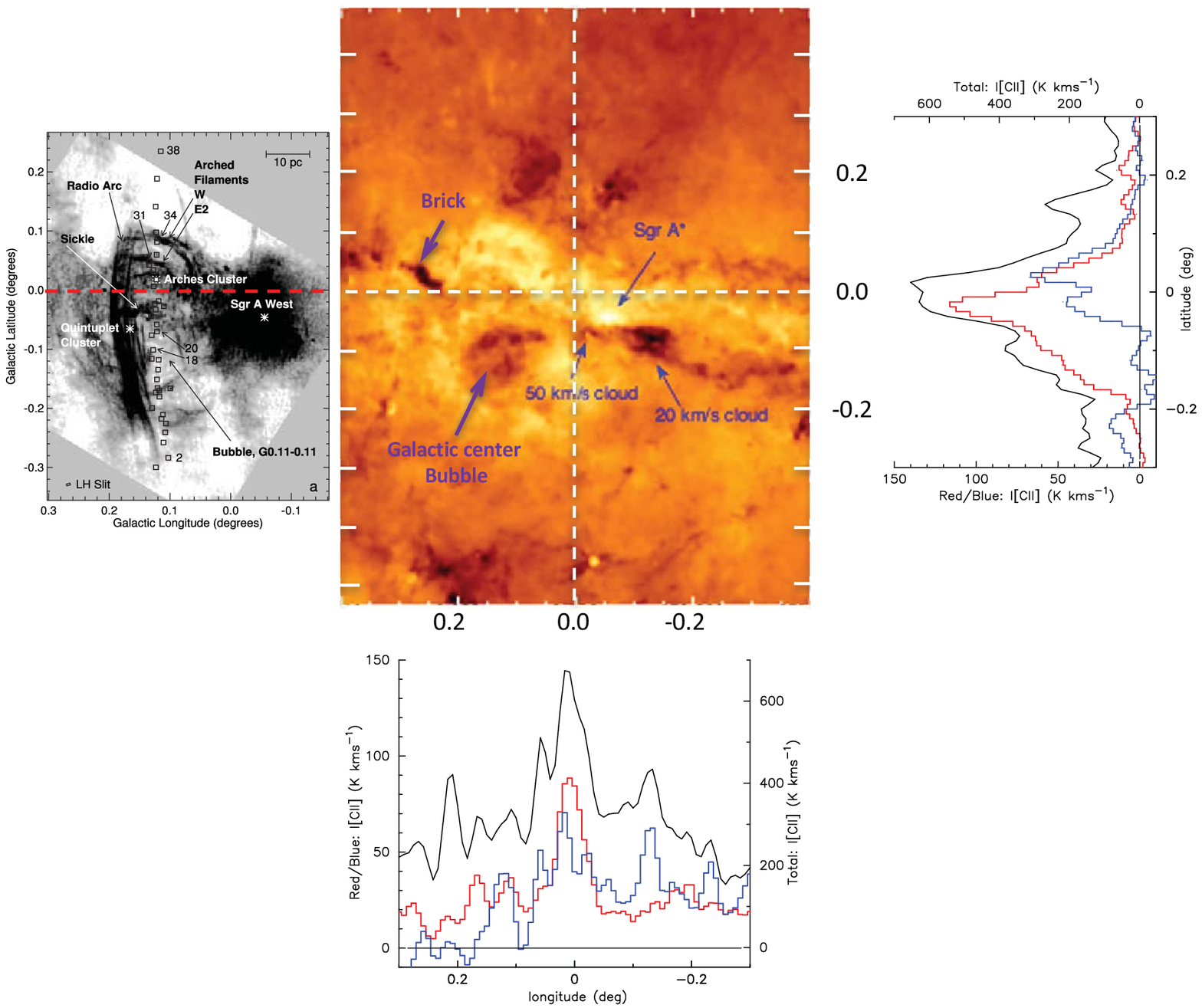}
        \caption{The center image is the  dust temperature distribution of the inner CMZ derived from {\it Herschel} HI-GAL 70--350 \micron images \citep{Molinari2011}.  It has a doughnut shaped distribution as seen in projection marking a Warm Dust Bubble \citep{Molinari2011,Molinari2014}. On the left  is  the radio continuum map generated by \cite{Simpson2007} based on the data of \cite{YusefZadeh1987a} of the central portion of the Bubble (the hole in the doughnut) containing the Radio Arc and Arched Filaments. The \cii  integrated intensity, I(\ciino) (K \kmsno), strip scans as a function of longitude and latitude are plotted to the bottom and the right, respectively, and are aligned with the temperature map of the inner CMZ.  These plots show  the velocity--integrated \cii intensity profiles as functions of longitude or  latitude: the black lines represent the total intensity in the velocity range V$_{LSR}$ = $-$200 to +140 \kmsno, the red and blue lines represent $I$(\ciino) integrated emission in the high--velocity red and blue wings from V$_{LSR}$ = +80 to +130 \kms and $-$80 to $-$130 \kmsno, respectively.}
         \label{fig:fig3}      
  \end{figure*}

In Figure~\ref{fig:fig3} we show the central portion of the dust temperature map of the CMZ from Figure 3 in \cite{Molinari2011}, which was derived from the {\it Herschel} Hi-GAL 70--350 \micron survey, (see also Figure 5 in \cite{Molinari2014}).  The Galactic Center Bubble is surrounded by warm dust with a doughnut-like temperature distribution seen in projection,  located to the left and below the Galactic Center.  The central portion of the doughnut is an ionized gas bubble centered at about (0\fdg11,--0\fdg11).  The dust temperature map shown in Figure~\ref{fig:fig3} covers about 0\fdg76 in longitude and 0\fdg 96 in latitude.  The white dashed lines are the \cii OTF strip scans. The  \cii longitudinal strip scan goes through the warm dust doughnut, while the latitudinal strip scan appears to be at the edge of the warm dust doughnut.   Also shown is a radio continuum map of the innermost portion of the CMZ \citep{YusefZadeh1987a,Simpson2007} revealing the Radio Arc, the Arched Filaments, and the clusters of massive stars associated with the Arches and Quintuplet Clusters.   The red dashed lines designate the \cii OTF strip scans through this region and it can be seen that the \cii longitude strip scan goes through the Arched Filaments and passes near the Arches Cluster. 
 
The  Arched Filaments have mostly negative velocities in the range  $-$70 to +15 \kms  \citep{Lang2001,Simpson2007}.  As can be seen in Figure~\ref{fig:fig2}(a),  \cii is detected along the line-of-sight near the base of the Arched Filaments, $l$$\sim$0\fdg06 \cite[see Figures 1 in][]{Lang2001,Lang2002} with a velocity range within those of the filaments.  It is also detected along $l$$\sim$0\fdg16, that might also be associated with the Arched Filaments, but is more likely associated with the molecular clouds with $V_{LSR}\sim$ --20 to --40 \kmsno.  There is also  \cii emission at $l\sim$0\fdg20 and with velocities $\sim$+20 to +30 \kms that is probably associated with the molecular cloud M0.20-0.033 \citep{Serabyn1994}.  

 In Figure~\ref{fig:fig3}  we plot the total integrated intensity $I$(\ciino) = $\int T_{mb}dv$ (K \kmsno) over the velocity range $-$200 to +140 \kms (black lines) as a function of longitude and latitude, bottom and right, respectively.  We also plot the integrated \cii intensity, $I$(\ciino), as a function of longitude and latitude for two different velocity ranges. The V$_{LSR}$ velocity range is  80 to 130 \kms (red) away from and  $-$80 to $-$130 \kms (blue) towards  the sun.  The $l$-- and $b$-- scan  intensity profiles   in Figure~\ref{fig:fig3}  indicate significantly large \cii emission present at the lower velocities  ($|V_{LSR}| <$ 80 \kmsno) in comparison to that in the high velocity wings ($|V_{LSR}| >$ 80 \kmsno). The strip scan in latitude shows a shift in $b$ in the peak \cii emission between the blue and redshifted components. A likely explanation for the latitude offset between these two velocity regimes is the tilt in the CMZ \cite[][and references therein]{morris1996}. 
 
           \begin{figure*}[!ht]
             \centering
       \includegraphics[width=18cm]{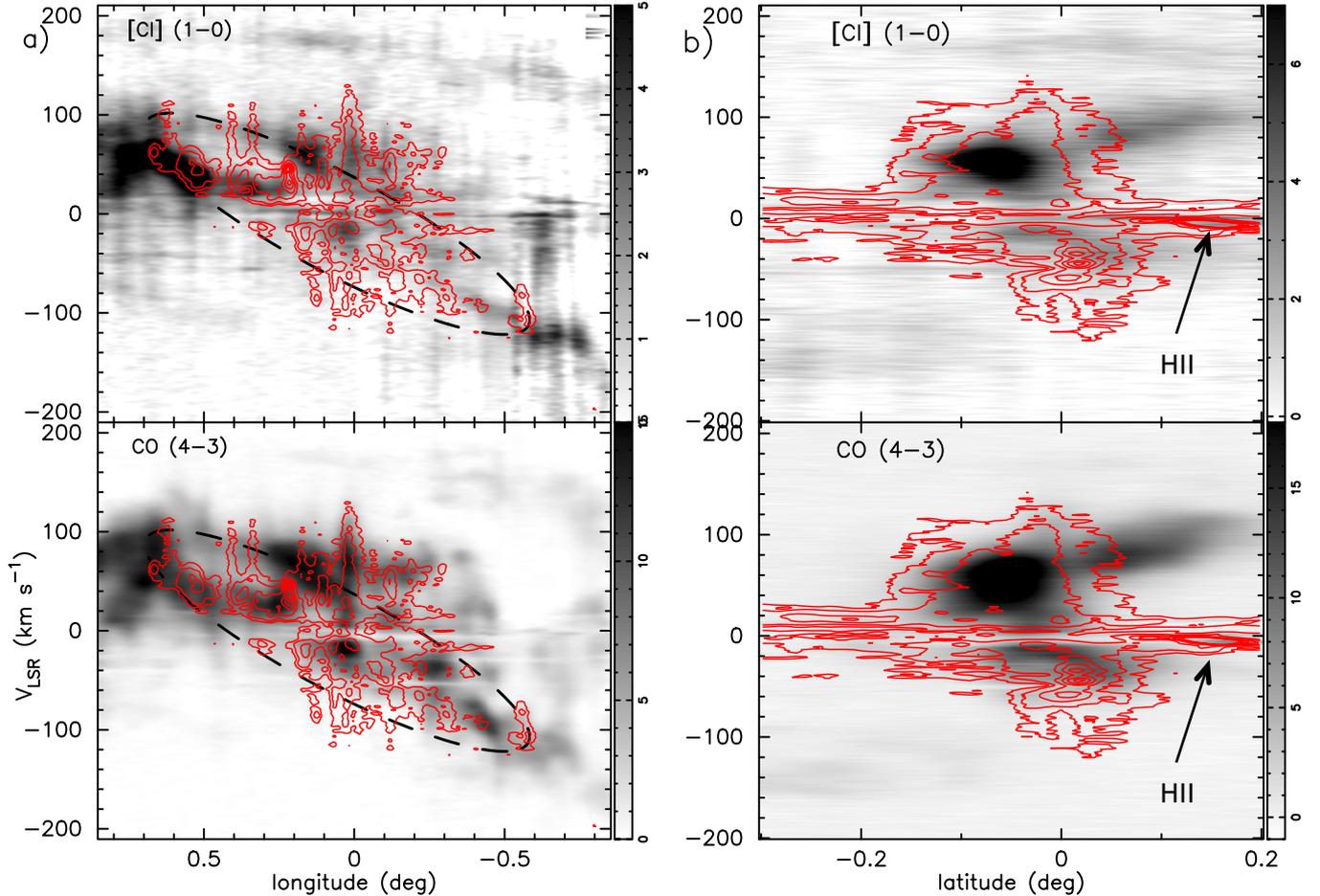}
\caption{Comparison of the \cii position--velocity strip scans with the AST/RO observations of \ci and CO(4--3) \citep{Martin2004}. (a) The \cii longitude--velocity strip scan and (b) the latitude--velocity strip scan.  The \ci and CO maps are shown in gray scale, where the grey scale bars on the right hand side indicate the main beam temperatures, $T_{mb}$(K); the corresponding \cii emission maps are overlaid as contours.  The \cii contour  levels (red) are $T_{mb}$(K) = 1, 2, 4, 6, 8, and 10.  In the $l$--$V$ maps the dashed line  is roughly outlines the  \cii emission at $b$=0\fdg0 for  the elliptical ring model of \cite{Molinari2011} (see also Figure~\ref{fig:fig2}).  This ellipse is used only to delineate the $l$--$V$ space where the open orbit gas streams can be found, and is not intended as a model of the gas dynamics. The arrows in panel (b) indicate a local \hii source (see Section~\ref{sec:S17})}
 \label{fig:fig4}
\end{figure*}
 
In Figure~\ref{fig:fig4} we compare the gas in the CMZ traced by \C+ with their neutral cloud components, C and CO, using the  AST/RO maps of \ci $^2P_1$$\rightarrow$$^2P_0$ and CO(4--3) \citep{Martin2004}. The AST/RO spatial maps cover a longitude range  $-$1\fdg 3 to +2\fdg 0, larger than the GOT C+ OTF strip scan, but were limited in latitude to $-$0\fdg 3 to +0\fdg 2, smaller than the GOT C+ strip scan.  The AST/RO maps have an angular resolution $\sim$120\arcsecno, which is about 50\% larger along the scan direction than the effective resolution in the GOT C+ strip scans.  The AST/RO spectra were resampled to match the velocity resolution of 2 \kms in the GOT C+ \cii OTF scans.  CO(4--3) and \ci in the AST/RO maps are strong at $b <$0\deg \cite[see Figure 3 in][]{Martin2004} and relatively weak at $b >$0\degno. The left panels show the $l$--$V$ maps and the right panels the $b$--$V$ maps. The features in these strip scans will be discussed in more detail in Section~\ref{sec:discussion}.

The $l$--$V$ maps (left panels) of all three tracers show the same general trend in velocity with longitude, but the velocity dispersion near $l$ = 0\deg is larger in \cii than in \ci  or CO(4--3).  A strong peak in the \cii longitude strip scan occurs at $l$$\sim$0\fdg 25, corresponding to the location of the Brick. In addition there are some notable anti-correlations within the $l$--$V$ maps, some features have strong \cii but weak or no \ci and/or CO, for example at $l$$\sim$0\fdg 25, and vice-versa.   In the latitude--velocity maps (right panels) \ci and CO(4--3) are generally correlated and, again, the \cii emission is weak where \ci and CO(4--3) are strong, and vice-versa. We also note a strong \cii feature at $b \sim$0\fdg15 which is a local \hii region (see Section~\ref{sec:S17}).  Finally, the two local peaks in \cii centered at $l$$\sim$0\fdg06  and 0\fdg18 with $V_{LSR}$$\sim$ --20 to --40 \kms appear associated with the Arched Filaments and Radio Arc filaments, respectively, but have little \ci or CO(4--3) associated with them.
        
The absence of a strong correlation between \cii and \ci and between \cii and CO(4--3) seen in many positions in Figure~\ref{fig:fig4} indicates that a significant fraction of the \cii emission in the CMZ is tracing a highly ionized gas component.   Furthermore, there is a trend for the highly ionized gas to appear closer to the Galactic Center than further out from it. To explore this trend further and to look for any differences in the source of \cii emission along the CMZ, we evaluated the global relative intensity distribution along the Galactic  longitude strip scan of \cii with respect to CO(4--3) and \ci by computing their average integrated intensities, $I$=$<$($\int{Tdv}$)$>$, in bins of 0\fdg5 width. The  average  intensity in each bin was calculated using a velocity range of $\pm$130 \kms but excluding the range $-$10 to +10 \kms which is contaminated with emission and absorption from local gas.  The results are summarized in Table~\ref{tab:Table_1}. Note that the \ciino/CO(4--3)  ratio at the center bin, within $\pm$0\fdg25 of (0\fdg0,0\fdg0), is about 2.3 times that on either side. Furthermore, the \ciino/\ci ratio at the center bin is also about 3 times that on either side.  Thus the results in Table~\ref{tab:Table_1} clearly show that there  is a distinct  excess in \cii over CO(4--3) and \ci within $\pm$0\fdg25 of the Galactic Center.  Unlike CO or \cino,  the \cii emission can arise  in both the neutral medium, excited by collisions with H$_2$ molecules, and in the ionized gas, excited by collisions with electrons.  Therefore, by comparison with the regions outside the central bins,  we can conclude that the \cii emission within $\pm$0\fdg25 of the Galactic Center has a significant contribution due to electron excitation of highly ionized gas.  This result is further corroborated by the detection of \nii in the Galactic Center \citep{Goldsmith2015,Garcia2016,Langer2016}.

\begin{table*}[!htbp]																		
\caption{Longitudinal \cii versus CO(4--3) and \ci in the CMZ}
\label{tab:Table_1}													
		
\begin{tabular}{cccccc}
\hline	
 Longitude & $<I$(\ciino)$>^a$  & $<I$(CO(4--3)$^b$))$>^a$ & $<I$(\cino)$^b$$>^a$  & $<I$(\ciino)$>$/$<I$(CO(4--3))$>$ &  $<I$(\ciino)$>$/$<I$(\cino)$>$   \\
  Range & (K \kmsno)  & (K \kmsno) & (K \kmsno)   &   &     \\
  \hline
  \hline 
0\fdg 75 to 0\fdg 25& 123 & 1012 &  345 & 0.12 & 0.36   \\
0\fdg 25 to -0\fdg 25& 313 & 1013 &  269 & 0.31 & 1.16   \\
-0\fdg 25 to -0\fdg 75& 67 & 570 &  172 & 0.12 & 0.38   \\
\hline	
\end{tabular}
\\	
a)  Average intensity calculated over the velocity range $\pm$130 \kms but excluding the emission in the range $\pm$10 \kms where it is most likely to be contaminated with emission from local gas.  b) AST/RO observations from \cite{Martin2004}.
 \end{table*}												
                
\subsection{\cii  spectra}
\label{sec:CII_NII}

\begin{figure*}[!ht]
 \centering
              \includegraphics[width=18.4cm]{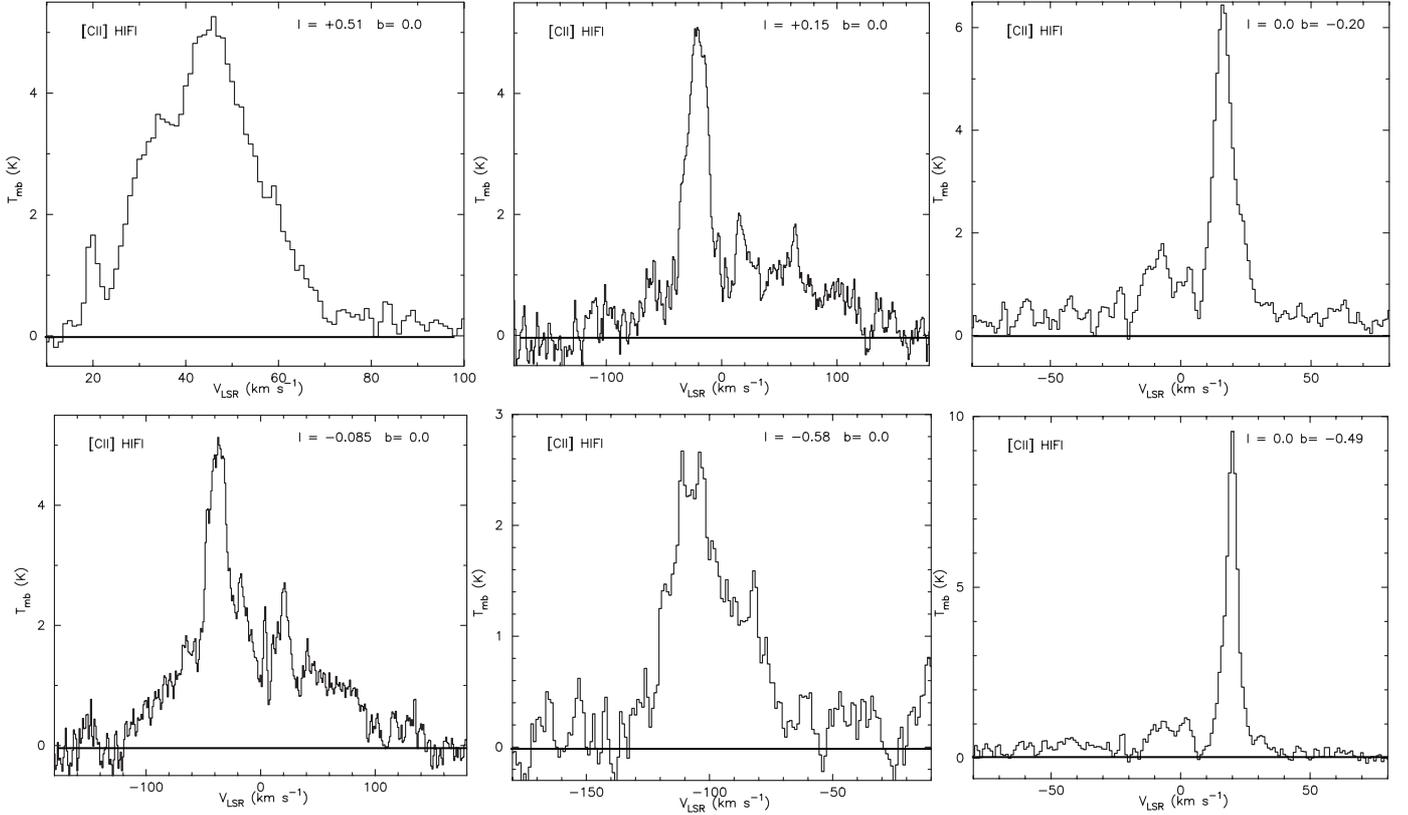}
      \caption{The GOT C+ \cii OTF spectra at  six positions where (0\fdg 51,0\fdg0) and ($-$0\fdg 58,0\fdg0) are at the edges of the open orbiting gas stream, ($-$0\fdg 085,0\fdg0) is near Sgr A, and  (0\fdg0,$-$0\fdg20) and (0\fdg0,$-$0\fdg49) show the narrowing of the \cii emission below the plane.}
         \label{fig:fig5}         
  \end{figure*}

In Figure~\ref{fig:fig5} we show six \cii spectra extracted from the OTF observations.  The spectra at (0\fdg51,0\fdg0) and ($-$0\fdg 58,0\fdg0) are at the hypothetical edges of the open orbit gas streams, ($-$0\fdg085,0\fdg0) is $\sim$8 pc outside Sgr A, and (0\fdg0,$-$0\fdg20) and (0\fdg0,$-$0\fdg49) show that \cii emission is significantly narrower below the plane. Within each spectrum there are narrow features with linewidths on the order of 10 to 15 \kmsno, along with broad but weaker \ciino.  In addition, there is evidence of absorption by local gas near 0 \kms in the innermost spectrum at ($-$0\fdg085,0\fdg0).
 
\begin{figure}
 \centering
             \includegraphics[width=7cm]{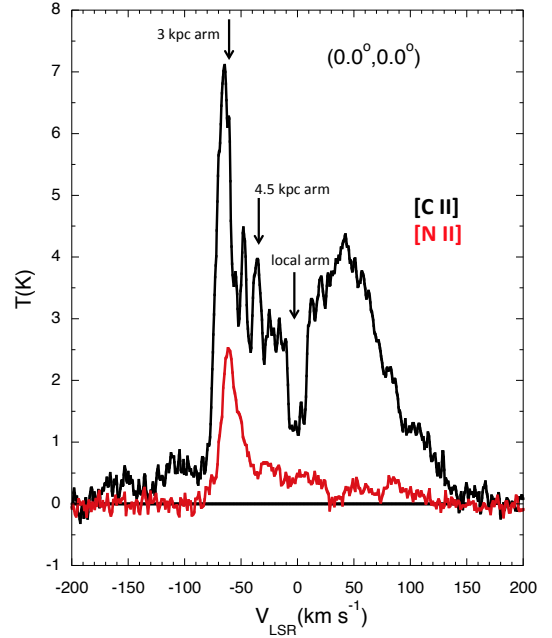}
      \caption{Main beam temperature versus velocity for \cii (black) and \nii (red) spectra towards (0\degno,0\degno) taken with HIFI  (Langer et al. 2016). The velocities of the foreground spiral arms are labeled.}
         \label{fig:fig6}       
 \end{figure}

At ($l$,$b$)=(0\degno,0\degno), in addition to a pointed observation of \cii from GOT C+, we also have a spectrum of the \nii 205 \micron fine structure line  taken from a {\it Herschel} HIFI survey of ten GOT C+ lines of sight \cite[][Langer et al. 2016]{Goldsmith2015}.  The angular resolution for the \nii spectra is 16\arcsec compared to 12\arcsec for \ciino.  In Figure~\ref{fig:fig6} the \nii spectrum is overlaid on the \cii spectrum\footnote{We reproduce the \cii and \nii spectra from \cite{Langer2016} here because the OTF strip scans show the spatial extent of the \cii component associated with \nii in the 3-kpc arm.}. It can be seen that there is emission across most of the velocity range of the \ciino, but only one strong component.   The line of sight to the Galactic Center intercepts a number of spiral arms.  We have indicated three features likely associated with the 3-kpc and 4.5-kpc arms, and local gas.  The strong peak in \nii aligns with the \cii peak towards the 3-kpc arm and likely most of the emission arises there.  A Gaussian fit to this \nii spectral feature yields $V_{LSR}$$\sim$ $-$60 \kms and  a FWHM line width $\sim$17 \kmsno.  The strongest \cii feature has very similar line parameters, $V_{LSR}$$\sim$ --64 \kms and FWHM  $\sim$19 \kmsno.   The couple of \kms shift in $V_{LSR}$ between \cii and \nii is characteristic of the emission arising from different regions across the spiral arm profile \citep{Velusamy2015}. The remaining \nii is broad low level emission arising from the CMZ and it is difficult to extract any distinct spectral lines with Gaussian-like shape.  The 4.5 kpc arm aligns with a peak in \ciino. The absorption by the local arm and local foreground gas at $V_{LSR}$$\sim$ $-$10 to +10 \kmsno, which is possibly in the Riegel-Crutcher cold cloud \citep{Riegel1972}, and which is very evident in \ciino, is not apparent in the \nii spectrum.  The N$^+$ abundance is a factor of a few less than \C+ and probably has too low an opacity to absorb significantly the \nii emission from the CMZ. 

The \nii line can be used to separate the contributions of \cii from highly ionized gas and neutral gas (PDRs and CO-dark H$_2$).  The \cii intensity from the fully ionized gas, $I_{ion}$(\ciino), is approximately proportional to the \nii intensity over a wide range of temperature and electron density, $I_{ion}$(\ciino)$\sim$0.9(C$^+$/N$^+$)$I$(\niino), where (C$^+$/N$^+$) is the carbon to nitrogen elemental abundance ratio \citep{Langer2016}.  The \cii intensity from the neutral gas,  $I_{neutral}$(\ciino), is just the difference of the observed total  \cii intensity,  $I_{tot}$(\ciino), minus the predicted $I_{ion}$(\ciino) derived from \niino. We calculated the fraction of emission of \cii from the ionized gas along  (0\degno,0\degno) using the spectra shown in Figure~\ref{fig:fig6}.  However, to derive just the contribution from the CMZ we have to exclude the emission from the 3-kpc arm which dominates the total \nii intensity.  We find that the predicted  $I_{ion}$(\ciino) derived from \nii is about 30\% of the total  \cii emission, $I_{tot}$(\ciino), as derived from the \cii spectrum, and the remaining 70\% comes from PDRs and CO-dark \h2 regions.  This result holds only for this one path through the CMZ because we do not have \nii spectra along the remaining positions in the two strip scans.  

\section{Discussion}
\label{sec:discussion}

In this section we discuss the properties of the following CMZ features traced by or near to the \cii OTF strip scans: the open orbit gas streams, the Brick, the Radio Arc, the base of the Arched Filament, the Warm Dust Bubble, Sgr B2, and an \hii region likely associated with the local source S17.   

 \subsection{Open orbit gas streams}
 \label{sec:Streams}
 
\cite{Molinari2011} found that {\it Herschel} far-infrared images of the CMZ revealed cold dense gas tracing out a roughly 200 pc size  $\infty$--shaped image, as seen in projection, centered slightly below $(l,b)$ = (0\degno,0\degno).  This feature corresponds well with the bow-like streams seen in CS by \cite{Tsuboi1999}.  \cite{Molinari2011} used the CS(1--0) spectral line maps and, assuming that this line traces  dense cores in the clouds forming this feature, they assigned average velocities to the points along the $\infty$-shape. They proposed that the shape and velocity structure could be modeled roughly as an orbiting elliptical ring with an oscillating warp in the $z$--direction.  They made a best fit to the CS data and found an ellipse with major and minor axes of $\sim$100 pc and $\sim$60 pc, respectively, projected on the plane of the Galaxy (see their Figure 5), an orbital velocity of 80 \kms (for simplicity assumed constant), and that the minor axis is oriented 40\deg with respect to the line of sight to the Sun.  The ellipse is assumed to lie 15 pc below the plane and to have a vertical frequency twice the orbital frequency.  The model parameters, as noted by \cite{Molinari2011}, need to be used with caution given the oversimplification of the model and that the assigned CS(1--0) velocities are derived by eye.  Nonetheless, the observed velocities are within $\pm$ 25 \kms of the model velocities.

However, as pointed out by \cite{Kruijssen2015} it is physically impossible to have a closed orbit in an extended gravitational potential. Instead they have replaced the single--orbit elliptical model of \cite{Molinari2011} by considering gas streams moving in an open orbit in the gravitational potential of the CMZ. \cite{Henshaw2016b,Henshaw2016a} have expanded on this model. They find a best fit to the observed velocity field if they assume four  streams (see Figure 2 in \cite{Henshaw2016a}). Furthermore, their orbital velocities vary along the orbit and are larger (up to 100 -- 200 \kmsno), than those in the elliptical ring model. Their best fit yields major and major and minor axes 121 and 59 pc, respectively, and an orbital velocity at apocenter and pericenter of 101 and 207 \kmsno, respectively.  

To see whether the \cii $l$--$V$  data encompasses the open orbit gas streams we derived the \cii velocities as a function of longitude for $b$=0\deg using the elliptical model of \cite{Molinari2011} to find the envelope of these features in \ciino. This elliptical model also allows us to constrain the limits on orbital velocity. (Note that our use of the \cite{Molinari2011} model to define the approximate $(l,b,V)$ extent of dense molecular gas does not imply that the gas is on closed orbits.) The solid red line ellipse in Figure~\ref{fig:fig2}(a) denotes the elliptical ring in longitude--velocity  from \cite{Molinari2011} calculated for $b$=0\degno. Our fit to the \cii data is shown as a red-dashed ellipse and, in order to  accommodate the emission at  $l$=$\pm$0\fdg6, requires an increase in orbital velocity to $\sim$100 \kms and orbit size of the major-axis to about 120 pc, as well as a slight change in the minor axis orientation to the line of sight to the sun \cite[see Figure 5][]{Molinari2011} to $\sim$50\degno.   In general the \cii emission follows the trend in the elliptical model, but the overall extent in longitude in \cii is larger than that derived by \cite{Molinari2011} from CS(1--0), as is evident at the limits of the ellipse at $l\simeq$$\pm$0\fdg6. In summary, the orbital velocity derived by \cite{Molinari2011} is too small to accommodate all the \cii emission.  

\begin{figure*}[!ht]
 \centering
                   \includegraphics[width=17.cm]{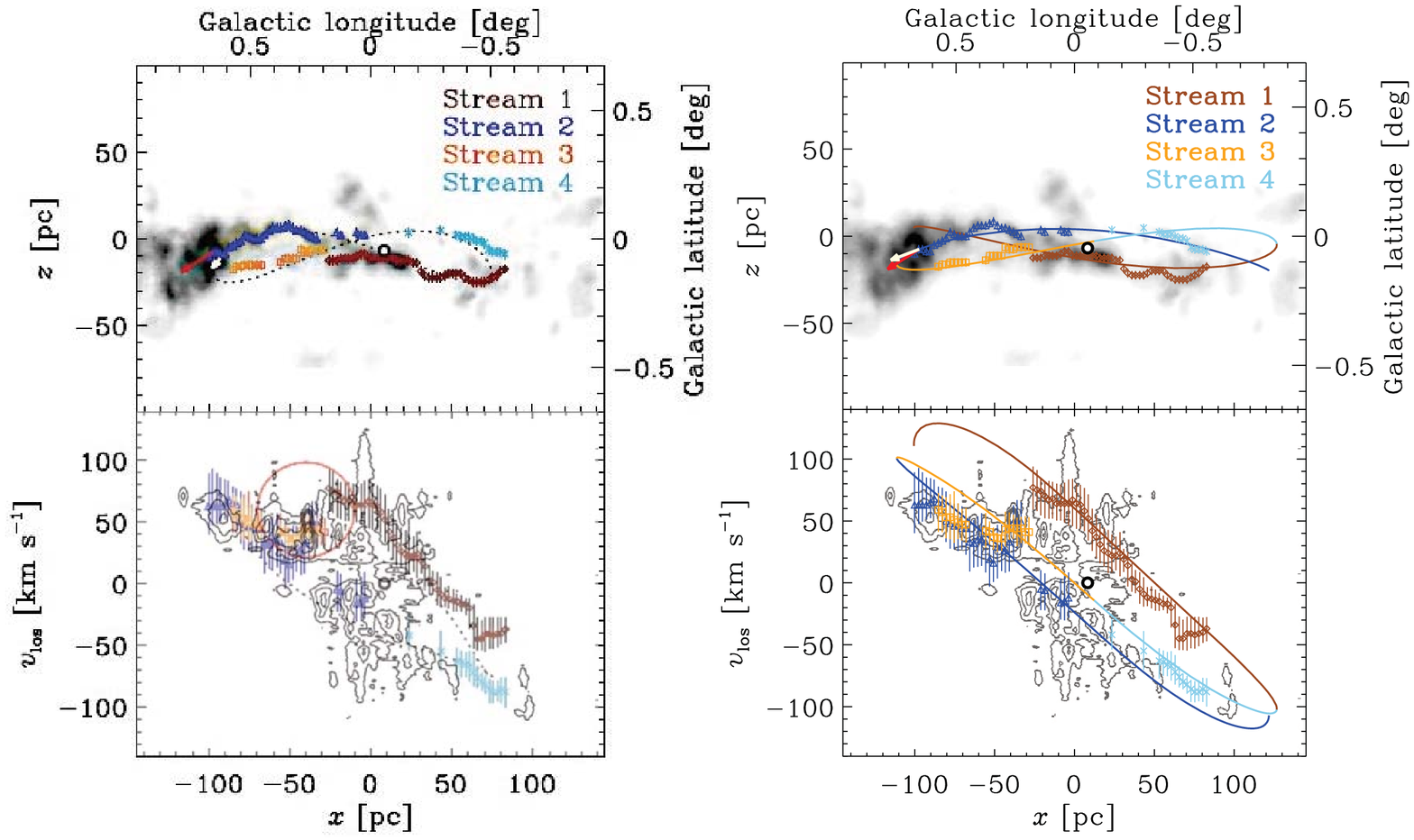}
        \caption{The GOT C+ \cii ($l$--$V$) map compared with the \cite{Molinari2011} orbital model and with the four open stream orbital model of \cite{Kruijssen2015}.   (left panels): ($l$--$b$) and ($l$--$V$) maps in which the dotted line indicates   the \cite{Molinari2011} orbital model (reproduced from Figure 2 in \cite{Kruijssen2015}).      The grey--scale in the top panel is the integrated intensity of NH$_3$(1,1) emission that traces the dense gas. The symbols with error bars represent the phase--space
information extracted by \cite{Kruijssen2015} from NH$_3$(1,1).  The small open black circle denotes Sgr A$^*$.  (The circle in the left panel is in the original figure and indicates a feature in position--velocity space discussed by  \cite{Kruijssen2015}). (right panels)  ($l$--$b$) and ($l$--$V$) maps in which the   solid open arcs  indicate  the four open gas streams and the legend labels them by color  (reproduced from Figure 4 in \cite{Kruijssen2015}). The GOT C+ OTF scan at $b$ =0\deg passes through streams 2 and 4, and passes closest to stream 3 near $l \sim$0\fdg2 and to stream 4 near $l \sim$0\fdg05.  In the bottom panel the GOT C+ \cii longitudinal--velocity contour plots are overlayed on the $l$--$V$ plots of the four stream open orbit model from \cite{Kruijssen2015}. }
         \label{fig:fig7}        
  \end{figure*}

Our  elliptical fit to the \cii emission is in better agreement with the \cite{Kruijssen2015} and \cite{Henshaw2016a}  open orbit gas stream models. To get a clearer picture of the relationship of \cii emission to the open orbit gas streams, we compare the GOT C+ \cii ($l$--$V$) map with the with the four open stream orbital model of \cite{Kruijssen2015} and the  \cite{Molinari2011} elliptical model.  Figure~\ref{fig:fig7} shows the ($l$--$b$) and ($l$--$V$) maps in \cite{Kruijssen2015} reproduced from their Figure 2 (left panel in Figure~\ref{fig:fig7}) and Figure 4 (right panel in Figure~\ref{fig:fig7}).  The dotted line in the left panel indicates   the \cite{Molinari2011} elliptical orbital model.   The solid open arcs in the right panel show the four open gas streams and the legend labels them by color. The grey--scale in the top panels  is the integrated intensity of NH$_3$(1,1) emission that traces the dense gas. The symbols with error bars represent the phase--space information extracted by \cite{Kruijssen2015} from NH$_3$(1,1) -- see their paper for details. The GOT C+ OTF scan at $b$ =0\deg passes through streams 2 and 4, and passes closest to stream 3 near $l\sim$0\fdg2 and to stream 4 near $l\sim$0\fdg05.  

In the bottom panel the GOT C+ \cii longitudinal--velocity contour plots are overlayed on the $l$--$V$ plots of the four stream open orbit model from \cite{Kruijssen2015}. The \cii ($l$--$V$) maps overlap both the elliptical and open stream models, but, as discussed above,  the outermost ($l$--$b$) \cii emission fits better with the open stream model. The \cii $l$--$V$ emission correlates well with streams 1 and 2, and a portion of stream 3, while the correlation with stream 4 is unclear due to the weakness of the \cii and overlap with stream 2.  However, the association of \cii emission  with the dense gas in streams 2 and 3 is perhaps not surprising as both of them pass through the \cii scan at $b$ = 0\degno, however the association with the dense gas in stream 1 (and perhaps stream 4), which pass a few pc below the \cii strip scan, suggests that there is lower density gas in a PDR, CO-dark H$_2$, or ionized region associated with the dense gas streams.  \cite{Longmore2013a} and \cite{Henshaw2016a}  propose that the gas along a stream represents a time sequence in the formation and evolution of star--forming dense clouds.  If there is a time sequence then perhaps some of the \cii in in the ($l$--$V$) map is tracing the  CO-dark \h2 phase in early cloud formation or a late stage where it is being stripped away.    If so then maps of the \cii emission may be able to tell us something about the dynamics of the formation of the clouds in the gas streams. 

One of the more interesting aspects of the orbiting gas streams is the presence of a corrugated velocity field with several regularly spaced  features detected over  $l\approx$ $-$0\fdg1 to $-$0\fdg7 \citep{Henshaw2016b}, whose velocity maxima correlate with massive and compact clouds.  \cite{Henshaw2016b} suggest that these features illustrate the mechanism of cloud formation in the inner CMZ and, in particular, highlight the regions where gas flow 'traffic jams' seed cloud formation.  The evidence for the ripples and condensations is derived from high density gas tracers such as N$_2$H$^+$, but direct evidence for the flows could come from spectral probes that  trace the lower density outlying gas surrounding the cores.  The \cii longitudinal strip scan also cuts across or passes close to most of the region in which \cite{Henshaw2016b} detect the ripples in velocity. 

\begin{figure}
 \centering
   \includegraphics[width=9cm]{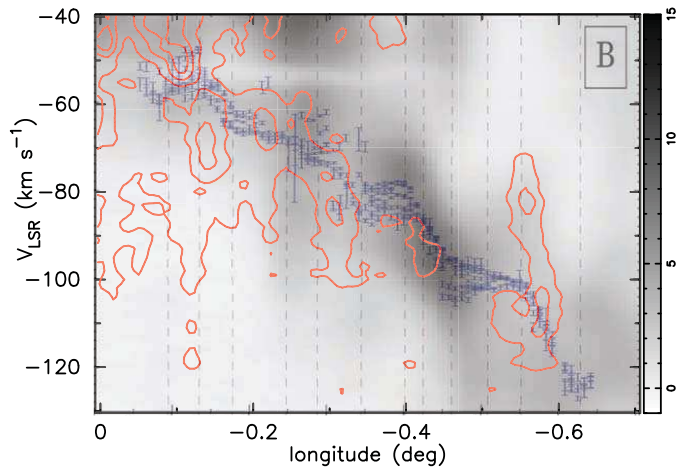}
        \caption{ Expanded view of the $l$--$V$ maps in Figure~\ref{fig:fig4} consisting of  \cii contours (red) and  the CO(4--3)  grey-scale image covering  $l$ = $-$0\fdg7 to 0\fdg0 and $V_{LSR}$ = $-$130 \kms to $-$40 \kmsno.  The \cii coutours are $T_{mb}$ = 1, 2, 3, 4, 5, and 6 K, and the CO(4--3) $T_{mb}$  values are shown in the colour bar on the right.   We have superimposed the centroid velocities of the dense cores presented in Figure 2b from the paper by \cite{Henshaw2016b}.  (The vertical dashed lines are in the original figure from \cite{Henshaw2016b} and identify the correspondence between the gas condensations and the extremes in velocity field amplitude.) It can be seen that the $l$--$V$ trends of \cii and CO(4--3) correlate well with those of the dense cores traced by N$_2$H$^+$.}
         \label{fig:fig8}   
 \end{figure}

To examine the relationship of \cii emission to the corrugated velocity field we show in Figure~\ref{fig:fig8} a blow-up of the \cii and CO(4--3) $l$--$V$ maps in Figure~\ref{fig:fig4} (bottom-left panel) covering the region studied in \cite{Henshaw2016b} in N$_2$H$^+$.  The region extends over $l$ = $-$0\fdg7 to 0\fdg0 and $V_{LSR}$ = $-$130 \kms  to $-$40 \kmsno.  For comparison to the dense gas tracers we overlay our \cii $l$--$V$ maps on Figure 2b from \cite{Henshaw2016b},  which is a longitude--velocity maps of the centroid velocities of the N$_2$H$^+$ spectral components. The N$_2$H$^+$ $l$--$V$ plot lies close to several \cii local peaks  that appear to have a corrugated velocity structure in $l$--$V$ space.  In addition, the CO(4--3) color image and \ci color image (not shown) appear to follow the gas streams.  Thus the \ciino, \cino, and CO(4--3) emissions are tracing the gas clouds in the open orbiting stream in this region.  This relationship is to be expected, as the neutral C and CO layers lie close to, and in the case of CO, overlap the dense cores in cloud chemical models, and \cii lies in the associated PDRs, CO-dark H$_2$ gas clouds, and ionized boundary layers. 

The widespread distribution of \cii as revealed by the large distribution in velocity and the   blending of \cii components makes it difficult to tell whether there is a systematic shift in velocity among the diffuse and dense gas tracers.  Only spectrally resolved $l$--$b$ maps of \cii of the orbiting gas stream can reveal whether there is evidence for the dynamical flows  responsible for cloud formation in this region. To derive the absolute relationship between the \cii velocity field and that of N$_2$H$^+$ is beyond the scope of this paper, given the limits inherent in the longitudinal  strip scan and the blending of many features in each \cii spectrum.  

\subsection{G0.253+0.016 (the Brick)}
\label{sec:Brick}

G0.253+0.016, also known as the Brick, is an infrared-dark cloud at $l \sim$0\fdg25 centered slightly above $b$ =0\degno.  There is  little evidence of massive star formation in the Brick despite its high mass, $\sim$(1 -- 2)$\times$10$^5$ \Ms, and high local and average densities, $>$10$^5$  and $>$10$^4$   cm$^{-3}$, respectively \citep{Lis1994,Longmore2012,Rathborne2014,Mills2015}.   The detection of one H$_2$O maser and several compact radio sources is consistent with the formation of perhaps at most a few early B-stars \citep{Longmore2014,Mills2015}.   \cite{Pillai2015} derive the orientation and strength of the magnetic field in G0.253+0.016 from dust polarization maps.  They find that the magnetic field is highly ordered and lies in an arch along the length of G0.253+0.016 and is sufficiently strong to support the cloud against collapse. \cite{Longmore2013a} suggest that this cloud was tidally compressed perpendicular to its orbit and stretched along its orbit after passing through pericentre at its minimum in the Galactic gravitational potential.  The \cii emission towards the Brick is strong; in fact it has some of the  brightest peaks in the CMZ strip scans, as can be seen in Figures~\ref{fig:fig2}(a)  and \ref{fig:fig4}.

 \begin{figure*}[!ht]
 \centering
                \includegraphics[width=15.4cm]{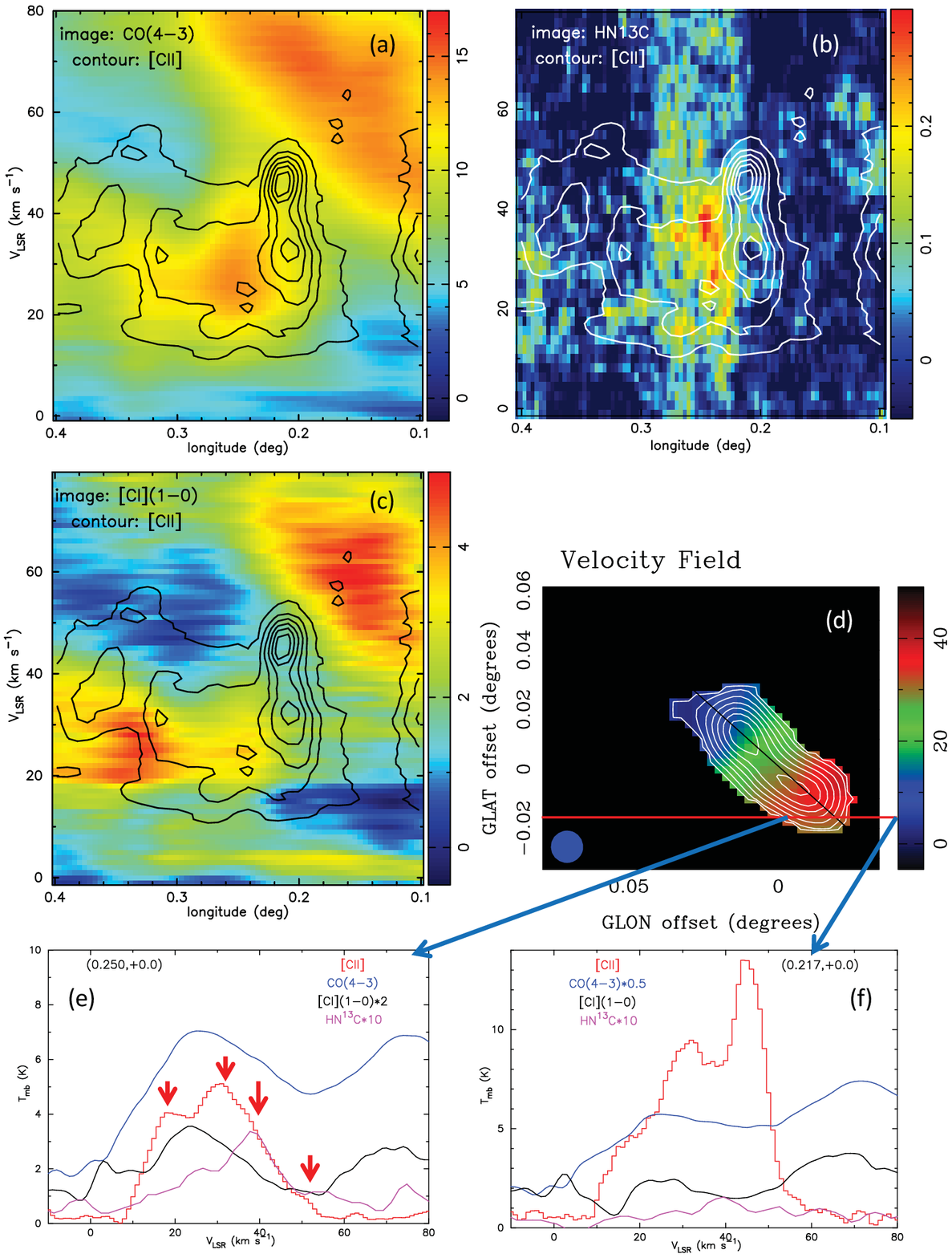}
                             \caption{Detailed view of the \cii longitude--velocity contour plot of $T_{mb}(K)$ associated with the Brick plotted over a map of  (a) CO(4--3) emission \citep{Martin2004}, (b)  HN$^{13}$C Mopra data (this paper), and (c) \ci  emission \citep{Martin2004}. The color bars give the values of $T_{mb}$(K) in the images; \cii contour levels are 2, 4, 6, 8, and 10 K, and the peak is 13.5 K.   (d) The velocity field and HNCO intensity of the Brick from Figure 6 in \cite{Rathborne2014} where the color bar on the right is the velocity field in \kms and the intensity is shown by the contours.  This plot is in offsets from the nominal center of the Brick, ($l$,$b$) =(0\fdg253,0\fdg016). The horizontal red line designates $b$ =0\deg and the blue arrows indicate the locations of the spectra shown in panels (e) and (f) which compare the GOT C+ \cii with the AST/RO \ci and CO(4--3) spectra \citep{Martin2004}, and the Mopra  HN$^{13}$C spectra (this paper) at  $l$ = 0\fdg250 and 0\fdg21667, respectively.  The red arrows in (e) indicate the four velocity components identified by \cite{Henshaw2016a} in the vicinity of G0.253+0.016 at $V_{LSR} \sim$7, 21, 38, and 70 \kms from dense gas tracers.  The HN$^{13}$C peak at $\sim$38.4 \kms is the feature arising from the Brick at this position.  The intensities  are given as$T_{mb}$(K). }       
         \label{fig:fig9}      
  \end{figure*}

G0.253+0.016 has been well studied by \cite{Rathborne2014} using a suite of optically thin and thick spectral line map data of high density gas tracers, and dust continuum emission in the mid-infrared and 450 \micronno.   In Figure~\ref{fig:fig9}(d) we show the integrated HNCO intensity distribution from Figure 6 in \cite{Rathborne2014} that traces out the dense core and the velocity field along the Brick.  Our \cii longitudinal strip scan is offset from the main core of the Brick as indicated by the horizontal red line at $b$ =0\degno.  It can be seen that the \cii spectral line data presented here pass through an edge of the Brick where $V_{LSR} \sim$ 30 to 40 \kmsno.  In contrast to the high dipole molecules, CO, \cino, and \cii trace the less dense outer envelopes of G0.253+0.016, and  probes an important spatial regime where dynamical effects (such as gas compression and stripping) due the orbital streaming  motion of the clouds with respect to Galactic center are likely to be pronounced. Here we present the results of the \cii emission  and velocity structure and an analysis of the \h2 molecular gas seen by \ciino, \cino, and CO emissions. 

In Figure~\ref{fig:fig9}  we compare the  \cii $l$--$V$ scans from $l$ = 0\fdg1 to 0\fdg4 at $b$ =0\fdg0, to three other gas tracers: CO(4--3), \cino, and HN$^{13}$C.  Figure~\ref{fig:fig9}(a)  shows the \cii contours on top of a color image of the AST/RO CO(4--3) emission \citep{Martin2004}, where the CO(4--3) emission associated with G0.253+0.016 is strongest at $l \sim$0\fdg25 over a velocity range $V_{LSR} \sim$20 to 40 \kmsno.  The CO(4--3) emission over $l \sim$0\fdg1 to 0\fdg3 at velocities 50 to 80 \kmsno is not associated with G0.253+0.016 and likely lies far from it \cite[see footnote 15][]{Rathborne2014}.  In addition, there are two very bright \cii features that peak at $l \sim$0\fdg21.   
 
Figure~\ref{fig:fig9}(c)  shows the \cii $l$--$V$ emission contours plotted over a color image of the \ci emission.  \ci is much weaker than CO(4--3) across the Brick and is strongest at $l \sim$0\fdg32, near the edge of the \cii, and weak at $\sim$0\fdg21, where \cii is stronger.   The \cii peak at $V_{LSR} \sim$30 \kms likely originates from the Brick's PDR or an \hii region.

Figure~\ref{fig:fig9}(b) shows the \cii $l$--$V$ emission contours plotted over a color image of the HN$^{13}$C emission.  The HN$^{13}$C data are from the Mopra\footnote{The data are from the Mopra radio telescope, a part of the Australia Telescope National Facility which is funded by the Commonwealth of Australia for operation as a National Facility managed by CSIRO. The University of New South Wales (UNSW) digital filter bank (the UNSW-MOPS) used for the observations with Mopra was provided with support from the Australian Research Council (ARC), UNSW, Sydney and Monash Universities, as well as the CSIRO} survey \citep{Jones2012}. The  HN$^{13}$C emission, while much weaker than \ciino, \cino, and CO(4--3), peaks at $l \sim$0\fdg24 and $V_{LSR} \sim$ 38 \kmsno.  It falls off sharply at $l \sim$0\fdg22 and this boundary marks the transition from the dense core to the outer envelope traced by \ciino.  Note also that the peak in HN$^{13}$C intensity is  distributed over $V_{LSR} \sim$35 to 45 \kmsno, whereas the adjacent \cii has a minimum.

There are two positions that we discuss in more detail, one is at ($l$,$b$)=(0\fdg250,0\fdg0),  hereafter G0.250+0.0, which is close to the nominal center of the dense core, and  the other ($l$,$b$) = (0\fdg217,0\fdg0), hereafter G0.217+0.0, is outside the edge of the dense core where \cii peaks. Here we discuss G0.250+0.0 and defer discussion of G0.217+0.0 to Section~\ref{sec:G0.217}.  Figure~\ref{fig:fig9}(e)  shows the \ciino, \cino, CO(4--3), and HN$^{13}$C spectra at $l$ = 0\fdg25, which is about 2.5 pc from the region associated with the nominal center of G0.253+0.016, and which is also the source of strong emission in the dense gas tracers.  It can be seen that there is emission by these tracers across the velocity range 10 to 50 \kmsno.  \cite{Henshaw2016a} has identified four velocity components in the region of G0.253+0.016 at $V_{LSR} \sim$7, 21, 38, and 70 \kmsno, only one of which, at 38 \kmsno, is associated with the Brick.

To understand better the velocity components in  the spectra arising from ($l$,$b$) =(0\fdg250,0\fdg0) we fit the line profiles with a multi-Gaussian function over the velocity range 0 to 60 \kmsno, which more than covers the velocity range of the Brick.  The two strongest gaussian components for HN$^{13}$C are at 26.0 and 38.4 \kmsno, and the latter is clearly seen in Figure~\ref{fig:fig9}(e). The component at 38.4 \kms is associated with the Brick and agrees with that derived by \cite{Henshaw2016a}. It is also closer to the $V_{LSR} \sim$36.5 \kms derived by \cite{Rathborne2014} at their position P4 in the Brick (see their Figure 7) which is about 1 pc away from (0\fdg250,0\fdg0).  The line parameters of the fits to \ciino, \cino, CO(4--3), and HN$^{13}$C for the Brick component are given in Table~\ref{tab:Table_2}. All four tracers have components associated with the Brick with $V_{LSR}$ within 1.0 \kms of each other, and full width half maximum (FWHM) linewidths $\sim$10.5 to 11.6 \kms with HN$^{13}$C being the narrowest line. The linewidths are in agreement with the intensity-weighted velocity dispersion derived by \cite{Rathborne2014} of $\sim$10 to 15 \kmsno.

In the open stream orbit models of \cite{Kruijssen2015} the core is compressed and collapses while the diffuse outer envelope is stripped away.  The three carbon species CO, C, and C$^+$  occupy largely different layers in a cloud, progressing from C$^+$ on the outside to   CO on the inside, with C mainly occupying a thin layer at the transition where all the main carbon species can coexist.  In principle, one could derive the relative motion of the different gas layers from the velocity peaks of the \ciino, \cino, CO(4--3), and HN$^{13}$C to test the formation models for G0.253+0.016 \cite[see discussion in][]{Rathborne2014}.  Adopting $V_{LSR} \sim$38.4 \kmsno, the peak in HN$^{13}$C, as the velocity of the core, then the C$^+$ and  C are blue shifted with respect to the core, which would be consistent with outflow of the envelope on the near side of the Brick, and infall on the far side.  Because the CO(4--3) is optically thick it almost certainly arises from the near side of the core however the velocity difference of 0.2 \kms is well within the error of the fits and no conclusion can be drawn about any dynamical velocity difference.   In addition, the blending of emission from many sources contributes to the spectra along this line of sight and even a difference of $\sim$1 \kms is  within the uncertainties of a multi-Gaussian fit to lines that are significantly blended. To make further progress we need to have $l$--$b$ \cii maps of this region to disentangle the velocity field.  

To calculate the column densities of CO, C, C$^+$, and HN$^{13}$C we use  the RADEX code \citep{vanderTak2007}, which is based on a large velocity gradient radiative transfer model, and the atomic and molecular data base of \cite{Schoier2005}.  Because \cii has only one transition and we do not have spectra for the upper \ci transition, $^2P_2$$\rightarrow$$^2P_1$,  we need to assume a temperature and H$_2$ density profile, as well as adopt a fractional abundance for carbon.  The first two inputs are guided by existing information from other sources on the temperatures and densities of the cloud.  The excitation results should be considered as a guide to possible conditions in the gas and not an exact solution. 

We assume a kinetic temperature in the CO(4--3) layer of 70K, similar to the gas temperature of 65 to 70 K  derived from H$_2$CO \citep{Ao2013}, and 100 K in the C and C$^+$ layers, typical of the gas temperatures in the envelope  \citep{Clark2013}.  For HN$^{13}$C we use the interior gas temperature of 65K.  At these high  kinetic temperatures the solutions are not that sensitive to uncertainties in the values chosen because they are comparable to or larger than the equivalent temperatures of the upper levels of CO(4--3), \cino, and \ciino, which  are  55.3, 23.6, and 91.3 K, respectively. For the density profile we assume that the density in the CO(4--3) layer is much less than the average core density of 2$\times$10$^4$ cm$^{-3}$, and adopt $n$(H$_2$) = 2$\times$10$^3$ cm$^{-3}$, which is much less than the critical density, $n_{cr}$(H$_2$) $\sim$4$\times$10$^4$ cm$^{-3}$, and thus the solutions are density sensitive.  For the \ci layer we also adopt $n$(H$_2$) = 2$\times$10$^3$ cm$^{-3}$, which is larger than the critical density $\sim$1.1$\times$10$^3$ cm$^{-3}$ and thus the \ci intensity is relatively insensitive to uncertainties in H$_2$ density.  In the \cii layer we adopt $n$(H$_2$) = 10$^3$ cm$^{-3}$ which is much less than the critical density $n_{cr}$(H$_2$) $\sim$4.5$\times$10$^3$ cm$^{-3}$. For the HN$^{13}$C layer we adopt 2$\times$10$^4$ cm$^{-3}$.  

To convert the column densities to a molecular \h2 column density adopt a fractional abundance of total carbon with respect to molecular hydrogen, $x$(C$_{tot}$) =2.8$\times$10$^{-4}$ as used in the models of \cite{Bertram2016}.  In most PDR models there is a very sharp transition from \C+ to CO as the UV radiation field is attenuated \cite[cf.][]{Bergin1997}, while neutral carbon is dominant in a thin layer \citep{Wolfire2010}.  For simplicity we assume that there is no mixing and the layers are either 100 percent ionized carbon, carbon monoxide, or neutral carbon, which is a reasonable approximation in the transition layer \cite[see][]{Wolfire2010}. We adopt a fractional abundance of total carbon with respect to molecular hydrogen, $x$(C$_{tot}$) =2.8$\times$10$^{-4}$ as used in the models of \cite{Bertram2016}.  In the case of HN$^{13}$C we adopt a fractional abundance of 3$\times$10$^{-9}$ and an isotope ratio $^{12}$C/$^{13}$C from \cite{Rathborne2014}, which is consistent with cloud model fractional abundances of  order (1 -- 6)$\times$10$^{-9}$ \citep{Bergin1997}.

The solutions are given in Table~\ref{tab:Table_2} where we list the assumed molecular hydrogen density, $n$(H$_2$), and kinetic temperature, $T_{kin}$, the solution to the column density of the tracer, $N$(X), the total H$_2$ column density, $N$(H$_2$), and the depth of the emission layer $L$ in pc. We also list the line opacity for each tracer and find that the \cii and \ci emissions are optically thin, while the CO(4--3) emission is optically thick. Because the CO(4--3) transition is optically thick and sensitive to the choice of density one must use this result with caution.  It can be seen that the \cii layer is the largest, extending over 0.8 pc, while the C layer is smallest, 0.15 pc, which is typical of PDR models.   The neutral carbon column density is relatively insensitive to the choice of local density, $n$(H$_2$), owing to the insensitivity of the upper level population to density for  densities $\ge$ the critical density.  The derived  thickness of the CO layer, which is only 0.35 pc, should be viewed with caution owing to uncertainty associated with the large opacity and sensitivity to choice of density. 

The HN$^{13}$C column density translates to $N$(HNC) $\sim$3.4$\times$10$^{14}$ cm$^{-2}$ at the edge compared to 10$^{15}$ cm$^{-2}$ derived by \cite{Rathborne2014} through the densest part of the core.   Our results at the edge along $b$=0\deg are consistent with the core values as the HN$^{13}$C intensities at the edge are about 10\% -20\% of those in the core. The column density of HN$^{13}$C corresponds to a depth of 1.8 pc. The total depth of the Brick at this location, as traced by the four carbon species is $L \sim$ 3.1 pc, as compared to 5.8 pc, twice the effective radius of the Brick \citep{Rathborne2014}.  Our estimate of the path length at the edge is consistent with the decrease expected from the core value.  

Most significantly,  the \cii emission is tracing some portion of the CO-dark gas surrounding the core.  Therefore,  it is likely that the \cii emission is probing a  thick layer of  \h2 without CO.  We cannot say whether this CO-dark \h2 gas is undergoing compression in a collapsing envelope or  being stripped away from a region where CO was dissociated.

\begin{table*}[!htbp]																		
\caption{Model column densities for G0.250+0.00 for velocity component with $V_{LSR}$ = 38 \kms}
\label{tab:Table_2}														
\begin{tabular}{llcccccccccc}
\hline	
  Species &Tracer & T$_{peak}^a$  & $V_{LSR}$&  FWHM &$n$(H$_2$)$^b$ & T$_{kin}^b$ & $\tau^c$ &$N$(X)  & $N$(H$_2$) & L \\
   &  & (K) & (\kmsno) & (\kmsno) & (cm$^{-3}$) & (K) & & (cm$^{-2}$)   & (cm$^{-2}$) &  (pc)   \\
  \hline
  \hline 
C$^+$ & \cii $^2P_{3/2}$$\rightarrow$$^2P_{1/2}$& 2.2 & 37.6 & 11.5 & 1$\times$10$^3$ &   100  & 0.34 & 7.2$\times$10$^{17}$ & 2.6$\times$10$^{21}$ & 0.83    \\
C & \ci $^2P_1$$\rightarrow$$^2P_0$ &  1.6 & 37.4 &  11.6 & 2$\times$10$^3$ &  100 & 0.02 & 2.5$\times$10$^{17}$  & 0.9$\times$10$^{21}$ & 0.15  \\
CO & CO(4--3) & 10.0 &  38.2 & 11.6 & 2$\times$10$^3$ & 70  &  6.3 & 6.0$\times$10$^{17}$  & 2.1$\times$10$^{21}$ & 0.35  \\
HN$^{13}$C & HN$^{13}$C(1--0) & 0.29 &  38.4 & 10.5 & 2$\times$10$^4$ & 65  &  0.65 & 1.7$\times$10$^{13}$  & 1.1$\times$10$^{23}$ & 1.8  \\
\hline	
\end{tabular}
\\	
(a)  Peak main beam temperature.  (b)  The densities and kinetic temperatures are inputs to the radiative transfer model. (c) Line opacity.							
 \end{table*} 
 
If the Brick is a site with little or no local star formation, especially lacking O stars, then the bright \cii emission from its edge at $\sim$ 32 \kms requires some other external source of ionization.   \cite{Mills2015} come to a similar conclusion from maps of widespread, but weak, radio continuum emission, tracing ionized gas near the Brick.  \cite{Mills2015} suggest that an O4-6 supergiant about 11 pc away (in projection) to the east is a plausible source of ionizing photons to explain the ionized gas giving rise to radio continuum emission.   Another possibility is that X-ray ionization is the source of the  high density hot ionized gas leading to intense \cii via electron excitation. Surveys of X-ray emission in the CMZ with {\it Chandra} \citep{Wang2002}, XMM  \citep{PontiMorris2015}, and {\it Suzaku} \citep{Koyama2007}, reveal a rich X-ray environment containing extended diffuse emission \citep{Koyama2007}, and over 9000 discrete sources \citep{Muno2009}.  As pointed out by \cite{Porquet2003} there is also a bright discrete source of X-rays, 1E1743.1+2843, located at $l$=0\fdg26 and $b$$\sim$0\fdg03, about 6 pc in projection from the Brick.  This source, which is the brightest Galactic Center X-ray source detected by XMM-Newton \citep{PontiMorris2015}, has an inferred X-ray luminosity $\sim$ 2$\times$10$^{36}$ erg s$^{-1}$ above 2 keV and is likely an accreting neutron star binary \citep{Porquet2003}.  However, one needs to know the luminosity down to 0.5 or 1 keV to evaluate its influence on the \cii region \citep{Langer2015X}. Unfortunately, X-rays in this range are absorbed by high column densities of intervening material and not measured directly, so we do not know the total relevant X-ray luminosity.

\subsection{\cii peak at G0.217+0.0}
\label{sec:G0.217}

The \cii emission at ($l$,$b$) =(0\fdg21667,0\fdg0), hereafter G0.217+0.0, has two of the strongest \cii emission peaks found in our strip scans. It is located close to or outside the edge of the dense core of the Brick (Figure~\ref{fig:fig9}(b)). In the position velocity maps in Figure~\ref{fig:fig9}(a) and (c) this feature is located between one dense core (the Brick) and a less dense core (seen in \ci and CO(4--3) at $V_{LSR} \sim$ 50  \kms but not in HN$^{13}$C. It is a relatively compact spatial feature in \cii that peaks at $l \sim$0\fdg21.   In Figure~\ref{fig:fig9}(f) we show the \ciino, \cino, CO(4--3),  and HN$^{13}$C spectra at this position. The velocity profile for \cii has two  peaks,  T$_{mb} \simeq$13.5 K at V$_{lsr} \sim$45 \kms and T$_{mb} \simeq$9.5 K at $\sim$30 \kmsno.  CO(4--3) is generally stronger than \cii at this location while \ci is much weaker. Neither CO(4--3) nor \ci show clear evidence of a double peak in velocity within the range of the \cii emission.  The HN$^{13}$C spectrum is very weak at these two positions, as might be expected outside the edge of the dense core.The multi-gaussian fit to the  \cii profile at this location yields  the following line parameters for the two prominent velocity features: $V_{LSR}$$\sim$ 31.1 and 44.5 \kmsno,  with $T_{mb} \sim$8.9 and 12.9 K, and FWHM $\sim$11.3 and 10.0 \kmsno, respectively.  The integrated intensities for these features are $I$(\ciino)$\simeq$ 101 and 130 K \kmsno. The strength of the \cii line and its spatial distribution in longitude suggest that  it likely arises from an ionized boundary layer or extended \hii region.  If we assume it comes entirely from the IBL then we can estimate the electron density in this region using an approach similar to that outlined in   \cite{Velusamy2012}.   Here we outline an approach to estimate $n(e)$. The \C+ column density as a function of the \cii intensity in the optically thin limit can be written \cite[][Equation 1]{Langer2015N},
  
\begin{equation}
N(C^+)=2.92\times10^{15}\big[1+\frac{e^{\Delta E/kT}}{2}(1+\frac{n_{cr}(e)}{n(e)})\big]I([C\,II])\,\,\rm{cm^{-2}\,}
\label{eqn:N_I_e}
\end{equation} 
 
\noindent where $I$(\ciino) is the intensity in K \kmsno, $\Delta E/k$ =91.21K is the energy needed to excite the $^2P_{3/2}$ level, $n(e)$ is the electron density, and $n_{cr}(e)$ is the corresponding  electron critical density.  The collision rate coefficient for electrons implies   $n_{cr}$(e)$\sim$50 cm$^{-3}$ at T$_{kin}$ = 8000K \citep{Goldsmith2012},  typical for highly ionized gas.   

If we assume that the density is uniform over the observed emission path length and that the gas is completely ionized then, $N$(C$^+$) = $n$(H$^+$)x(C$^+$)$L$ where $x$(C$^+$) is the fractional abundance of ionized carbon, and $L$ is the path length in cm. Substituting  $n$(e)=$n$(H$^+$) we have a quadratic equation relating $n$(e) to $I$(\ciino) and $L$,

\begin{equation}
\frac{n(e)^2} {[n(e)+ n_{cr}(e)/3]}= 4.4\times10^{15}\frac{I([C\,II])}{x(C^+)L(cm)}\,\,\rm{cm^{-3}}
\label{eqn:n(e)_I_L}
\end{equation}  

\noindent where we have set the factor exp($\Delta E/kT$) equal to 1, as $\Delta E/kT <<$1 in a hot ionized gas.  If we know the size of the emission region, $L$, we can solve  Equation~\ref{eqn:n(e)_I_L} for $n$(e) (or equivalently $n$(H$^+$)) as a function of $I$(\ciino) for a given $x(C^+)$ and $T_{kin}$.  The exact value of the kinetic temperature of the hot ionized gas is not critical to the solution because  $n_{cr}$(e) is a weak function of kinetic temperature, $\propto$T$_{kin}^{0.37}$ \citep{Goldsmith2012}.  

To estimate a scale size for the \cii emission region along the line of sight at (0\fdg217,0\fdg0) we assume that it is roughly the size of the longitudinal width of the \cii ridge in Figure~\ref{fig:fig9}, $\sim$0\fdg04, which at the distance to the Galactic Center is $\sim$5.8 pc.  Assuming this emission comes from  warm ionized gas with $T_k$ = 8000 K and a fractional abundance $x$(C$^+$)=2.8$\times$10$^{-4}$,  we find that $n$(e) $\sim$102 cm$^{-3}$ and 128 cm$^{-3}$ for the $V_{LSR}$$\sim$ 31.1 and 44.5 \kms features, respectively.  (If the emitting layer is more like a 2--D surface, then we have overestimated the path length and the density will be higher.)  
The density is consistent with those found from studies using the \nii far-infrared fine structure lines to determine $n$(e) in dense bright \hii regions \citep{Oberst2011} and in Sgr A \citep{Goldsmith2015}, but two to five times higher than those possibly characteristic of the ionized boundary layers (IBL) of clouds in the disk \citep{Goldsmith2015} and clouds at the edge of the CMZ \citep{Langer2015N}.  

\subsection{The Radio Arc Region}

The Radio Arc Region is a set of very prominent radio filament features near Sgr\,$\space$A consisting of the nonthermal filaments (NTF) of the Radio Arc, the thermal Radio Arches, and nonthermal Radio Threads, as can be seen in the  left panel of Figure~\ref{fig:fig3}, which is a radio continuum map presented in \cite{Simpson2007} and adapted from \cite{YusefZadeh1987a} (see also Figure 1 in \cite{Lang2001,Lang2002}).  The lines-of-sight of the GOT C+  longitudinal strip scan intercept the base of the  Arched Filaments over  $l \sim$0\fdg04 to $\sim$0\fdg08 and the NTF Radio Arc filaments from $l \sim$0\fdg15 to $\sim$0\fdg18.  

\begin{figure}
 \centering
            \includegraphics[width=8.5cm]{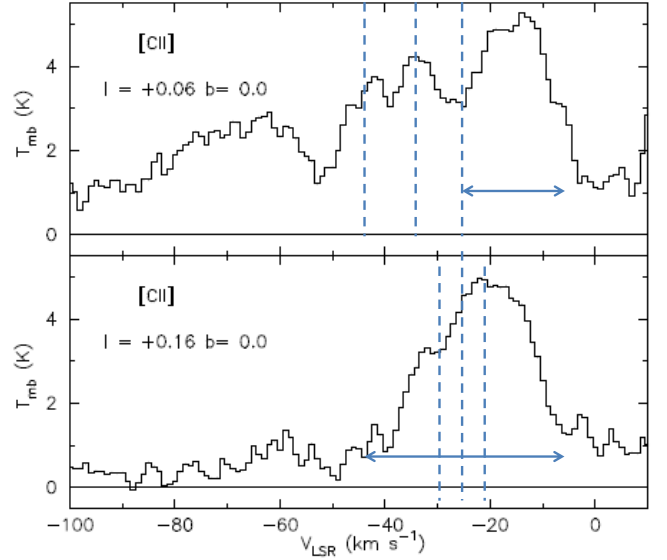}
        \caption{\cii spectra at two longitudes containing  negative velocity clouds underlying the arched radio filaments \citep{Lang2001}.   The dashed lines mark velocity features also observed in H92$\alpha$ radio recombination lines associated with the thermal (ionized gas) component in the nearby  radio arches.   The horizontal double--arrows mark the CS(2--1) velocity ranges for molecular clouds in the vicinity of  $l = $0\fdg06 \citep{Lang2002}  and $l = $0\fdg16 \citep{Serabyn1987}.}
         \label{fig:fig10}   
 \end{figure}

In Figure~\ref{fig:fig10} we mark the velocities of peak emission in the  H92$\alpha$ spectrum at three locations  from Figure 7 in \cite{Lang2001} which are closest to that of the \cii spectra  for  $l$ =0\fdg06 at $V_{LSR}$=$-$26 \kmsno, $-$43 \kmsno, and $-$34 \kms  (upper panel) in locations 15,16, and 17, respectively,   and for $l$ =0\fdg16 at $V_{LSR}$ =$-$21 \kmsno, $-$30 \kmsno, and $-$27  \kms (lower panel) in locations 1, 2,  and 8, respectively.  Note that all these locations are offset from $b$=0\deg  at slightly higher latitude ($\sim$2 to 3 arcmin) but their longitudes are within the effective beam size of the \cii map.  In the 8.3 GHz continuum map (Figure 7 in \cite{Lang2001}) all the arches near $l$ =0\fdg06 appear to extend down towards $b$ =0\deg where they seem to merge into a more diffuse structure.  The horizontal double--arrows mark the CS(2--1) velocity range for molecular clouds in the vicinity of  $l= $0\fdg06 \citep{Lang2002}  and $l$=0\fdg16 \citep{Serabyn1987}.   The broad \cii feature at $V_{LSR}$$\sim$--70 \kms in the spectrum for $l$ =0\fdg06 is associated with the open orbit gas streams (see Section~\ref{sec:Streams}).   

Detection of these velocity features in the \cii spectra at $l$ =0\fdg06 suggests significant contribution to the \cii emission from the ionized gas in the arches.  The close association of the two \cii velocities with the H92$\alpha$ spectra at locations 16 and 17 in \cite{Lang2001}, and the separation from the CS(2--1) molecular cloud emission, indicates that this \cii emission is likely tracing parts of the thermal \hi  filaments at $l$=0\fdg06.  The 8.3 GHz continuum  maps in this region along $b=$0\deg appear to trace the base of the arched thermal filaments, which are also seen in \cii at the velocities for the filaments.  
However, because the velocity of the feature at location 15 is within the range of the CS velocities we cannot tell whether it is coming from the base of this filament or the CS molecular cloud envelope.  
   
The three filaments in Figure 7 from \cite{Lang2001}  for $l$ =0\fdg16 that correspond to the velocities marked in Figure~\ref{fig:fig10} (lower panel) seem less pronounced in the region observed in \ciino.  Therefore, in contrast to the emission at $l \sim$0\fdg06, the  bright \cii emission at $l \sim$0\fdg16  in the velocity range $-$5  \kms to $-$45 \kms may arise from  an extension of the negative-velocity molecular cloud underlying the arched filaments, which shows CS emission in the velocity range  approximately $-$50 \kms to $\sim$0 \kms \cite[][]{Serabyn1987}. The bright \cii emission at $l \sim$0\fdg15 to 0\fdg18 occurs right at the interface between the dense clouds seen in CS and the NTFs, where the thermal radio emission is relatively weak. In addition, there is evidence for \nii emission from this region in the HEXGAL channel maps \cite[see Appendix F of][]{Garcia2015}.  Therefore, this emission might arise from  a dense PDR and/or the ionized boundary layer associated with the CS molecular clouds. 

At $l \sim$0\fdg06 the \cii emission from the thermal Arched Filaments  at $-$34 \kms and $-$43 \kms  probably arises from electron excitation of C$^+$.  For these components we can use the \cii emission to estimate the electron density associated with the base of the filaments following  an approach similar to that  for G0.217+0.0 in Section~\ref{sec:G0.217}. We assume that the filaments are completely ionized and at a kinetic temperature of 8000K.  (\cite{Lang2001} quote an ion temperature of 6900K, but the solutions to $n$(e) are relatively insensitive to this kinetic temperature difference, as discussed in Section~\ref{sec:Brick}). The radio filaments in the 8.3 GHz continuum maps near (0\fdg06,0\fdg0) are about 50\arcsec  wide or $\sim$2 pc in width along the  longitudinal strip scan. Thus the \cii emission in each filament is slightly beam diluted along the scan direction by about a factor of 1.7.  If we assume that the filament is cylindrically symmetric, then  the path length of \cii emission is the same as the width of the filament.  We can use Equation~\ref{eqn:n(e)_I_L} to solve for the electron density.  The \cii components at $-$34 \kms and $-$43 \kms have an integrated intensity, $I$(\ciino), of 47 and 35 K \kmsno, respectively.  Putting all these factors into  Equation~\ref{eqn:n(e)_I_L} yields $n$(e) $\sim$100 cm$^{-3}$ and 80 cm$^{-3}$ at $-$34 \kms and $-$3 \kmsno, respectively, at the base of the Arched Filaments.   These densities  are lower than the electron density derived using the 8.3 GHz radio continuum intensity at peak F   \citep{Lang2001}, $n$(e) $\sim$340 cm$^{-3}$, located about 1\arcmin away in $b$.  The lower $n$(e) at $b$ =0\deg is consistent with the weaker and more diffuse emission in the 8.3 GHz maps near the base of the thermal Arched Filaments.

\subsection{\cii cloud emission models}
\label{sec:model}

Recently \cite{Clark2013} and \cite{Bertram2016} presented numerical hydrodynamical simulations of CMZ clouds that include many physical processes, including thermal heating and cooling, chemical reaction networks, and radiative transfer models, with the goal of producing synthetic emission maps for comparison to observations.  In particular, \cite{Bertram2016} produce velocity integrated intensity maps for  \cii (see their Figure 4) for three different initial values of the ratio of the cloud's kinetic to potential energy, $\alpha$ = 0.5, 2.0, and 8.0. The peak intensities in the spatial-spatial maps $I$(\ciino) are of order 50 to 100 K \kms  regardless of $\alpha$, but the larger the value of $\alpha$ the larger the physical dimensions of the cloud. We cannot make a direct comparison of the GOT C+ observed \cii intensities to the model maps because we only have strip scans.  However, for the  \cii integrated intensities and the size of the clouds intercepted in the longitudinal strip scans, excluding the region near Sgr A, are in the range 50 to 110 K \kmsno, similar to the models in \cite{Bertram2016} $\alpha$ =0.5.  
     
\subsection{Warm Dust Bubble}

The Warm Dust Bubble, as seen in the dust temperature image in Figure~\ref{fig:fig3} \cite[Figure 3 in][]{Molinari2011}, has the appearance in projection of a doughnut.  \cite{Molinari2014} describe it as a very young superbubble with a roughly 15 pc radius containing the Arches and Quintuplet Clusters.  The exact nature of the energy source powering this bubble remains unclear.  The GOT C+ latitudinal  strip scan passes through the ring of  the warm dust doughnut while the longitudinal  strip scan is about halfway between the upper and lower edge of the ring. In Figure~\ref{fig:fig3} we also plot the integrated \cii intensity, $I$(\ciino) = $\int T_{mb}dv$ (K \kmsno), as a function of longitude (bottom) and latitude (right side) for two different LSR velocity ranges:  80 to 130 \kms away (red) and $-$80 to $-$130 \kms towards (blue) the sun, corresponding to the far and near sides of the bubble, respectively, assuming that the bubble is expanding. The strip scan in latitude shows a shift in $b$ in the peak \cii emission between the blue-- and red--shifted components. These two velocity ranges may be tracing the expansion of the gas associated with the dust bubble.  Although it is also possible that the latitude offset between these two velocity regimes is attributable to the tilt in the CMZ \cite[][and references therein]{morris1996}. 

 \subsection{Sgr B2}

Sgr B2 is  one of the most massive molecular clouds in the Galaxy and a well-known star-forming region.  The compact \hii regions, Sgr B2(N) at ($l$,$b$) $\sim$(0\fdg6773,$-$0\fdg02776) and Sgr B2(M) at ($l$,$b$) $\sim$(0\fdg6671,$-$0\fdg0350) \citep{Lis1994}, are two very active star forming cores with a $V_{LSR} \sim$60 to 70 \kmsno. They lie about 4 pc and 5 pc, respectively, below the \cii longitudinal strip scan.  The narrow beam size in latitude (12\arcsecno) in the \cii $l$--$V$ map precludes any detection from the Sgr B2(N) and Sgr B2(M) cores themselves, however it does sample the gas that may be influenced by their stellar activity.  As seen in Figures~\ref{fig:fig2}(a) and \ref{fig:fig4} at  ($l$,$b$)  $\sim$(0\fdg67,0\fdg0) there is strong CO(4--3) emission over a broad range in velocity, but  \cii emission is only over a narrow velocity range around $V_{LSR} \sim$65 \kmsno. In view of its proximity to the open orbit gas streams (dashed line) in Figure~\ref{fig:fig4} we discuss this feature in more detail.

\begin{figure}
 \centering
            \includegraphics[width=9cm]{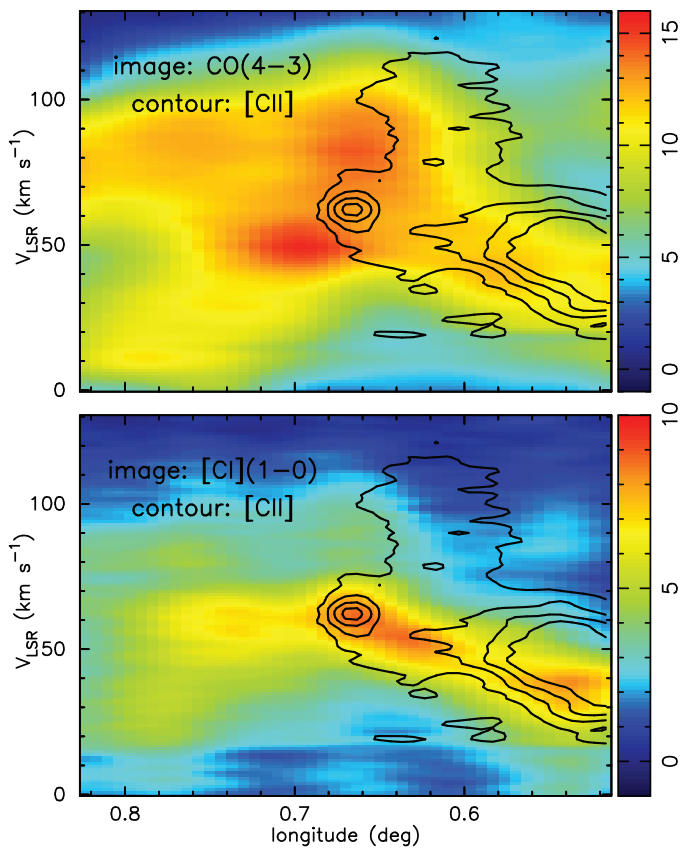}
        \caption{ (top) \cii $l$--$V$ contour map  including longitudes closest to Sgr B2 overlayed on an image of the AST/RO CO(4--3) emission. (bottom) The same for \cii overlayed on the \ci image.  The \cii contours are in T$_{mb}$(K) and have the values 0.75, 1.5, 2.25, and 3.0 K.  The values of T$_{mb}$(K) for the CO and \ci images are given by the color bars on the right of each panel. } 
         \label{fig:fig11}       
 \end{figure}

\begin{figure}
 \centering
            \includegraphics[width=7cm]{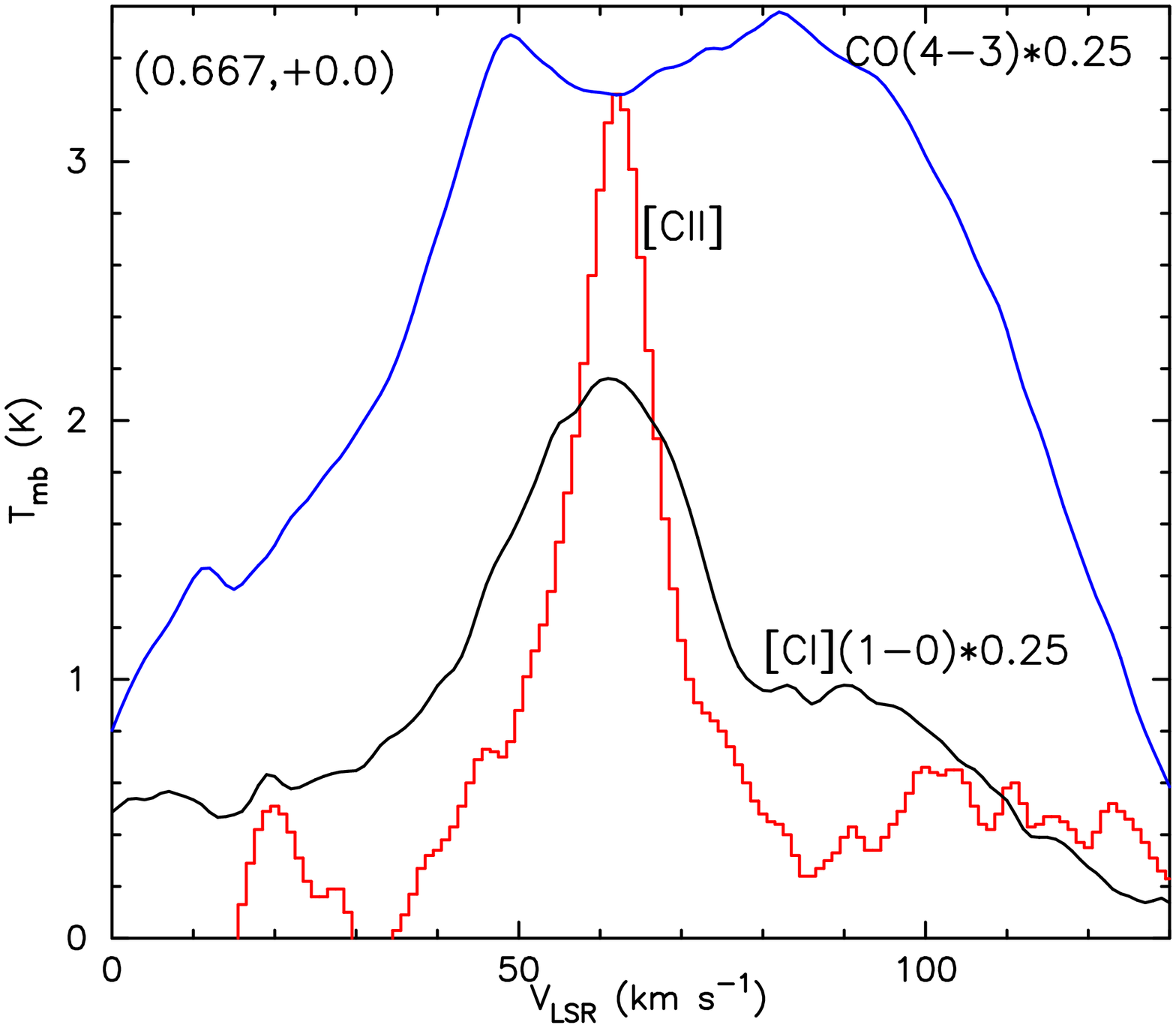}
        \caption{ \ciino, CO(4--3), and \ci spectra at the position of peak emission near Sgr B2,  ($l$,$b$) = (0\fdg667,0\fdg0), in the longitudinal scan. } 
         \label{fig:fig12}          
 \end{figure}

 In Figure~\ref{fig:fig11}  we show an expanded view of the \cii emission in this region compared to that of \ci and CO(4-3) and in Figure~\ref{fig:fig12}  we show their spectra at the peak, ($l,b$) = (0\fdg667,0\fdg0).   Near Sgr B2 the \cii spectrum peaks at $V_{LSR} \simeq$62 \kms with a linewidth FWHM $\simeq$14 \kmsno, and is much narrower than CO(4--3), which has broad emission over 100 \kmsno.  The dip in the CO spectrum at this velocity could be due either to the superposition of many CO sources or absorption by lower excitation CO.   In the models of \cite{Molinari2011} and \cite{Kruijssen2015} Sgr B2 is located at or near the tangent point of the streaming gas orbit so that many gas clouds would be expected to lie along this line-of-sight, but they would be spread over a range of distances.  Thus the broad CO line is sampling  many clouds along the line-of-sight in contrast to the \cii line which appears to be strong only at the gas having a velocity associated with Sgr B2. The \ci line also peaks at $V_{LSR} \simeq$62 \kms but is broader then \ciino, with a FWHM $\simeq$ 32 \kmsno. Most likely the \cino, as in the case of CO, is coming from a number of gas clouds along the line of sight.    In summary, while the bulk of the CO and fainter diffuse \cii is associated with the open orbit gas streams, the compact \cii is likely an ionized boundary layer or comes from  \hii regions associated with Sgr B2.  
  
  \subsection{\hii region S17}
  \label{sec:S17}

There is very strong compact \cii emission at $b \sim$0\fdg15 in the latitude--velocity plots at  V$_{LSR} \sim$ $-$5 \kms as seen in Figure~\ref{fig:fig2}(d)  and Figure~\ref{fig:fig4} (right panel). The peak temperature $T_{mb}$(\ciino) $\sim$35 K and the size of the emission region in $b$ is about 0\fdg4.  The relative narrowness of the \cii spectra, FWHM $\sim$4 to 5 \kmsno, is more typical of clouds in the disk than the CMZ.    The maps in Figure~\ref{fig:fig4} show that this region has weak \ci and CO(4--3) emissions.  We cannot be certain of the location of this \hii source because with V$_{LSR} \sim$ $-$5 \kms it could lie almost anywhere along the line of sight to the CMZ.  However, the narrowness of the \cii  spectrum and the fact that it lines up almost exactly with an \hii source in the WISE Catalog of Galactic \hii Regions  \citep{Anderson2014}, suggests it is a local source associated with the large \hii region designated S17 in the Sharpless catalog \citep{Sharpless1959}, rather than arising from the CMZ.

    


\section{Summary}
\label{sec:summary}

We report on the first high spectral resolution OTF strip scans of \cii (158 \micronno) across the CMZ taken with {\it Herschel} HIFI as part of the GOT C+ Open Time Key Programme.  A primary goal of this survey was to explore what types of features within the CMZ  can be detected in \cii and what can be learned about the kinematics and properties of the gas containing C$^+$. This survey consisted of two OTF strip scans, each 1\fdg6 long centered on (0\degno,0\degno), one in longitude and the other in latitude.  The effective spatial resolution of this survey was $\sim$80\arcsec along the scan and 12\arcsec perpendicular to the scan.  \cii was detected in a wide variety of CMZ features including G0.253+0.016 (the Brick), the open orbit streams of gas, the Radio Arc, the Arched Filaments, and the gas near Sgr A and Sgr\,$\space$B2.    

We  found that \cii traces portions of the several open orbit streams of cloud-forming gas \cite[cf.][]{Longmore2013a,Kruijssen2015} that stretch over 200 pc across the CMZ \citep{Tsuboi1999,Molinari2011}.  A fit to the spatial--velocity longitudinal cut of \cii through $b$ =0\deg yields orbital velocities up to 100 \kms and a major axis  $\sim$120 pc,  consistent with the open orbit stream models of \cite{Kruijssen2015}.  We also find that \ciino, \cino, and CO(4--3) may trace the corrugated velocity features of dense gas in the streams studied  by \cite{Henshaw2016b}, but we are unable to draw any conclusions about the relative motion of these outer cloud gas tracers with respect to those of the dense cores. 

\cii emission is detected across  the Brick. There are two strong emission velocity features at $l \sim$ 0\fdg21667 located outside the dense core and CO image of the Brick's molecular gas, one of which is associated with the Brick.  The other feature is associated with another cloud at higher velocities that has \ci and CO but no evidence of the dense gas tracer HN$^{13}$C.  Both \cii features  likely originate in an ionized boundary layer or extended \hii region.   We estimated the electron density in these two regions to be about 100 to 140 cm$^{-3}$ assuming the \cii arises from a highly ionized gas.  We also analyze the conditions where the \cii longitudinal strip scan intersects the edge of the Brick by fitting the line parameters and calculating column densities of \ciino, \cino, CO(4--3), and HN$^{13}$C at G0.250+0.0. We find that the HNC column density is about 10\% of the value through the core, and that C$^+$ forms an envelope around the molecular core about 0.8 pc in depth. 

We detect two prominent \cii features at  $l \sim$0\fdg06 at velocities, $V_{LSR} \sim$ $-$34 to $-$43 \kms corresponding to the Radio Arc near the base of the Arched Filament.  Assuming that the \cii from this region arises primarily  from fully ionized gas we estimate that the electron densities, $n$(e) $\sim$ 100 to 130 cm$^{-3}$ at the base of the thermal Arched Filaments. 

The OTF strip scans intersect the Warm Dust Bubble as seen in projection as a doughnut, identified in the {\it Herschel} HI-GAL infrared survey \citep{Molinari2014}.  \cii emission appears to trace this warm dust, which is not surprising as the UV which keeps the dust and gas warm also gives rise to \C+ and, therefore, \ciino. We do not have sufficient information to know for certain whether the source of \cii is from PDRs, CO-dark \h2, or fully ionized gas.  However, an analysis of the intensity ratio of \cii to CO(4--3) and \cino(1--0) indicates that a significant fraction of the \cii in the inner $l$ =$\pm$0\fdg 25 along $b$ =0\deg arises from ionized gas. The observed velocity structure in the \cii emission is consistent with the kinematics of an expanding bubble.  Finally, we detect only weak \cii emission in the strip scan that passes within 4 to 5 pc (in projection) from Sgr B, an active region of star formation with strong compact \hii sources. 

 In summary the GOT C+ \cii survey of the CMZ, although limited to two narrow OTF strip scans, shows that the bright far--IR 158 \micron \C+ line, when observed with high spatial and spectral resolution, brings out a wide range of emission features tracing a variety of physical conditions in the interstellar gas in the CMZ and near the Galactic Center.  The GOT C+ strip scans demonstrate the need for future large-scale, spectrally resolved \cii  $l$--$b$ maps,  fully sampled in in both longitude and latitude.  Such maps, as well as those of other ions, such as \niino, will be a vital and important template for understanding galactic nuclei.




\begin{acknowledgements}
We would like to thank Dr. David Teyssier for clarifications regarding the use of the {\it hebCorrection} tool.  We also thank an anonymous referee for numerous in depth comments and suggestions that improved the analysis and discussion significantly. 
This work was performed at the Jet Propulsion Laboratory, California Institute of Technology, under contract with the National Aeronautics and Space Administration.   U. S. Government sponsorship acknowledged.
\end{acknowledgements}



\bibliographystyle{aa}
\bibliography{aa_CMZ_CII_Survey_refs}
 

\end{document}